\documentclass{aa}

  \usepackage[latin1]{inputenc}
  \usepackage{graphicx}
  \usepackage{tabularx}
  \usepackage{natbib}
  \usepackage{amsmath,amssymb}

  \newcommand{\eg}{e.g.}
  \newcommand{\etal}{et al.}
  \newcommand{\privcom}{{\it private communication}}
  \newcommand{\num}{n$^\circ$}
  \newcommand{\papi}{paper~I}
  \newcommand{\papii}{paper~II}
  \newcommand{\sms}[1]{{\mbox{{\tiny #1}}}}
  \newcommand{\ddiff}{{\, \rm d}}
  \newcommand{\msol}{\mbox{M$_\odot$}}
  \newcommand{\lsol}{\mbox{L$_\odot$}}
  \newcommand{\zsol}{\mbox{Z$_\odot$}}
  \newcommand{\mic}{\mbox{$\mu$m}}
  \newcommand{\ngc}[1]{NGC$\,$#1}
  
  \newcommand{\mw}{Milky Way}
  \newcommand{\iizw}{II$\,$Zw$\,$40}
  \newcommand{\hen}{He$\,$2-10}
  \newcommand{\henalt}{ESO$\,$495-G21}
  \newcommand{\sbs}{SBS$\,$0335-052}
  \newcommand{\xxxdor}{30$\,$Doradus}
  \newcommand{\all}{\iizw, \hen\ and \ngc{1140}}
  \newcommand{\oo}{On-Off}
  
  \newcommand{\tauiram}{$\tau_{230}$}
  \newcommand{\nir}{near-IR}
  \newcommand{\mir}{mid-IR}
  \newcommand{\fir}{far-IR}
  \newcommand{\fuv}{far-UV}
  \newcommand{\sfr}{$\rm M_\odot\, yr^{-1}$}
  \newcommand{\hi}{H$\,${\sc i}}
  \newcommand{\hii}{H$\,${\sc ii}}
  \newcommand{\ha}{H$\alpha$}
  \newcommand{\neiii}{Ne$\,${\sc iii}}
  \newcommand{\neii}{Ne$\,${\sc ii}}
  \newcommand{\siv}{S$\,${\sc iv}}
  \newcommand{\ariii}{Ar$\,${\sc iii}}
  \newcommand{\peg}{P\'EGASE}
  \newcommand{\clo}{CLOUDY}
  \newcommand{\dbp}{DBP90}
  \newcommand{\cc}{\c{c}}

  \newcounter{textlistctr}
  \newcommand{\thetextlist}{, }
  \newcommand{\textlist}[1]
                 {\setcounter{textlistctr}{1}
                  \renewcommand{\thetextlist}
                  {{\it (\roman{textlistctr})}\stepcounter{textlistctr}}#1
                  }

\begin{document}

\graphicspath{{figures/}}


\title{ISM Properties in Low-Metallicity Environments}
\subtitle{III. The Dust Spectral Energy Distributions of \all}
\titlerunning{The Dust Spectral Energy Distributions of \all}

\author{Fr\'ed\'eric~Galliano\inst{1}\thanks{current address: 
           Infrared Astrophysics Branch, Code 685, 
           NASA Goddard Space Flight Center, Greenbelt, MD 20771, USA} \and 
        Suzanne~C.~Madden\inst{1} \and
        Anthony~P.~Jones\inst{2} \and 
        Christine~D.~Wilson\inst{3} \and
        Jean-Philippe~Bernard\inst{4}}
\authorrunning{F.~Galliano \etal\ }
\institute{Service d'Astrophysique, CEA/Saclay, L'Orme des Merisiers,
             91191 Gif sur Yvette, France \and
           Institut d'Astrophysique Spatiale (IAS),
             Universit\'e de Paris XI, 91405 Orsay, France \and
           Department of Physics and Astronomy, McMaster University,
             Hamilton, ON L8S 4M1, Canada \and
           Centre d'\'Etude Spatiale des Rayonnements (CESR), 31028 
             Toulouse, France}

\abstract{
We present new 450 and 850$\,\mic$ SCUBA data and 1.3$\,$mm MAMBO data
of the dwarf galaxies \all.  
Additional ISOCAM, IRAS as well as ground based data are used to construct 
the observed mid-infrared to millimeter spectral energy distribution of these 
galaxies.  
These spectral energy distributions are modeled in a self-consistent way, as
was achieved with \ngc{1569} \citep{galliano+03}, synthesizing both
the global stellar radiation field and the dust emission, with further
constraints provided by the photoionisation of the gas.  
Our study shows that low-metallicity galaxies have very different dust
properties compared to the Galaxy.  
Our main results are:
    \textlist{\thetextlist~a paucity of PAHs which are likely destroyed by the 
      hard penetrating radiation field, 
    \thetextlist~a very small ($\sim 3-4\,\rm nm$) average size of grains, 
      consistent with the fragmentation and erosion of dust particles by the 
      numerous shocks,
    \thetextlist~a significant millimetre excess in the dust spectral energy
      distribution which can be 
      explained by the presence of ubiquitous very cold dust 
      ($\rm T = 5-9\, K$) accounting for 40 to 80$\,\%$ of the total dust 
      mass, probably distributed in small clumps.}
We derive a range of gas-to-dust mass ratios between 300 and 2000,
larger than the Galactic values and dust-to-metals ratios of $1/30$ to
$1/2$.  
The modeled dust size distributions are used to synthesize an
extinction curve for each galaxy.  
The UV slopes of the extinction curves resemble that observed in some regions 
in the Large Magellanic Cloud. 
The 2175$\,$\AA\ bumps of the modeled extinction curves are weaker than that 
of the Galaxy, except in the case of \iizw\, where we are unable to accurately
constrain the 2175$\,$\AA\ bump carrier.
}

\date{Received November 16, 2004 / Accepted}

\offprints{galliano@avak.gsfc.nasa.gov}

\maketitle


\section{Introduction}

Since dust can either reveal or conceal star formation activity, it
plays an important role, often as an elusive villain, in the
interpretation of the cosmic star formation history.  
While characterisation of the detailed dust properties is required to
understand the attenuation of the starlight, and to determine the in
situ star formation properties, it is difficult to achieve this due to
the necessity of sampling a broad range of the dust spectral energy
distribution (SED).  
We have been carrying out studies of detailed SED modeling, with the goal of 
reproducing a self-consistant model for emission and extinction in galaxies, 
in order to explore the effects of different environments on the dust 
properties. We began our efforts with low metallicity environments.  
What has been gleaned to date indicates that low metallicity environments 
can, indeed, harbor non-negligible quantities of dust
\citep{thuan+99,plante+02,galliano+03}, the effects of which can not be 
ignored. 
Additionally, the physical properties of the dust do not resemble those of 
the Galaxy \citep{galliano+03,vanzi+04}.
These results call to question the validity of the seemingly-innocuous
assumption of Galactic-dust properties.  
All of these results have striking implications on the interpretation of 
extragalactic SEDs, and influence our view of the cosmic star formation 
history.

This is the third paper in a series of publications studying the dust 
properties in star-bursting low-metallicity dwarf galaxies.
The first paper \citep[hereafter \papi;][see \citealt{madden00}, for a 
preview of this study]{madden+05} was a \mir\ 
spectroscopic study of a sample of dwarf galaxies observed by ISOCAM.
The second paper \citep[hereafter \papii]{galliano+03} presents a detailed 
model of the dust SED of the dwarf \ngc{1569}.
Here, we present a study similar to \papii, applied to the three starbursting
dwarf galaxies \all. 
In this paper we pull together all that we have learned from in depth 
analyses of an assortment of low-metallicity environments.

\iizw\ is a blue compact dwarf galaxy (BCD), at a distance of 
$D\simeq 10\,$Mpc, often considered to be the prototypical \hii\ galaxy, its 
nucleus being dominated by one large \hii\ region of $\sim 0.5\,$kpc diameter.
Its bright optical core is located inside a large \hi\ envelope.
The two tails seen in optical and near-infrared 
\citep[{\eg}][]{cairos+01a,vanzi+96}, are the relics of the merger of two 
smaller galaxies \citep{brinks+88}.
The starburst could be very young.
The metallicity has been determined to be $Z\sim 1/6\;\zsol$ 
\citep{masegosa+94}.
This galaxy is \hi\ rich 
\citep[$M(\mbox{\hi})=4.4\times 10^8\,\msol$~;][]{vanzee+98}, 
while the molecular gas emission is very weak 
\citep[$M_\sms{mol} < 0.4\times 10^6\,\msol$~;][]{meier+01}.
The radio continuum emission is compact and mainly thermal free-free 
\citep{beck+02}.

\hen\ (\henalt) is a very bright southern BCD galaxy, located at a distance of 
$D\simeq 9\,$Mpc.
It was the first Wolf-Rayet galaxy to be identified \citep{allen+76}.
Its peculiarities among BCD galaxies is that its metallicity appears to be 
almost solar \citep{kobulnicky+99b} and it has a relatively large amount 
of molecular gas \citep{meier+01}.
The starburst is concentrated in two regions separated by a dust lane.
It contains several Super Star Clusters (SSCs) probing a burst of a few Myr 
\citep{johnson+00}.
This recent burst could have been triggered by the interaction with a 
molecular cloud \citep{kobulnicky+95}.
A radio map \citep{kobulnicky+99a} reveals several compact thermal 
continuum sources, indicating ultra-dense \hii\ regions.
\ha\ images show a large scale bipolar outflow of gas \citep{mendez+99}.

\ngc{1140} is an amorphous starburst galaxy, larger than \iizw\ and \hen\ 
with a larger reservoir of gas.
Its size is roughly $1.7'\times 0.9'$ \citep{buat+02}, at the distance of 
$D\simeq 20-25\,$Mpc.
The central \hii\ complex of \ngc{1140} has an \ha\ luminosity which is 40 
times higher than that of \xxxdor\ \citep{hunter+94a} and contains several 
SSCs which have  been studied in detail by \citet{hunter+94b} and 
\citet{degrijs+04}.
Its star formation rate is about 1$\,$\sfr.
\ngc{1140} may be in the final stages of a merger \citep{hunter+94a} and has
a small dwarf companion.
Among the four dwarf galaxies of our sample (\ngc{1569}, \all), it is the only
one which shows pronounced PAH emission at \mir\ wavelengths (\papi).

The modelling of the global SED of a dwarf galaxy, using actual dust models, 
has been presented by \citet{madden00}, \citet{galliano+02}, 
\citet{lisenfeld+02}, \citet{galliano+03}, \citet{takeuchi+03} and this study
\citep[see also][]{galliano05}.
\citet{takeuchi+03} have used a dust evolution model to fit the IR SED of \sbs.
\citet{lisenfeld+02} and \citet{galliano+03} (\papii) applied the 
\citet{desert+90} model to \ngc{1569}. 
\citet{galliano+03}, having sufficent observational constraints, were able to 
constrain the dust size distribution of different dust components for 
\ngc{1569}. 
Our conclusion in \papii\ was that the global IR SED in \ngc{1569} is composed 
of dust components with very different size distributions from our Galaxy and 
contrary to our Galaxy, is dominated by {\it small dust particles} 
($\lesssim 10\,$nm). 
Additionally, a submillimetre excess was found that could be attributed to a 
very cold dust component.

Studies on the ISM of dwarf galaxies, on smaller spatial scales have
also been conducted. 
\citet{bot+04} applied the \citet{desert+90} model to the diffuse emission of 
the Small Magellanic Cloud (SMC).
While they did not vary the size distribution, they took into account
the variations of the radiation field.  
They had difficulty fitting the $60\,\mic$ flux.  
Finally, \citet{plante+02} and \citet{vanzi+04} have modelled the dust SED 
around the SSCs in \sbs\ and \ngc{5253}, respectively.  
They used Dusty \citep{dusty} which solves the radiative transfer equations 
in a spherical environment, without considering the process of stochastic 
heating.  
In these cases, the size distributions of the local environments were found 
to be dominated by large grains.  
These differerent results obtained for global galactic scales versus more 
local regions around the SSCs is likely due to the destruction of small 
grains around SSCs, as is also observed in the vicinity of AGNs 
\citep{maiolino+01b,maiolino+01a}.

The paper is organised as follows.
Section~\ref{sec:obs} presents an overview of our new
observations and the data we obtained from the literature.
Section~\ref{sec:results} presents the modeled SEDs of \all\ and the 
consequences of the results on the dust properties.
We end with a summary and the conclusions in Sect.~\ref{sec:concl}, which ties
together Papers I, II and III.


\section{The observations}
\label{sec:obs}

Observed SEDs were constructed as completely as possible, using our ISOCAM 
data (\papi), our new 450 and 850$\,\mic$ JCMT (SCUBA) observations and our 
IRAM (MAMBO) observations presented here, as well as incorporating
data from the literature for various telescopes and wavelengths. 
The differences in beam sizes are not of great concern for this paper, since 
we are modeling the global SEDs here.


  \subsection{SCUBA images}
  \label{sec:scuba}

\begin{figure*}[htbp]
  \centering
  \begin{tabular}{cc}
    \includegraphics[width=0.5\linewidth]{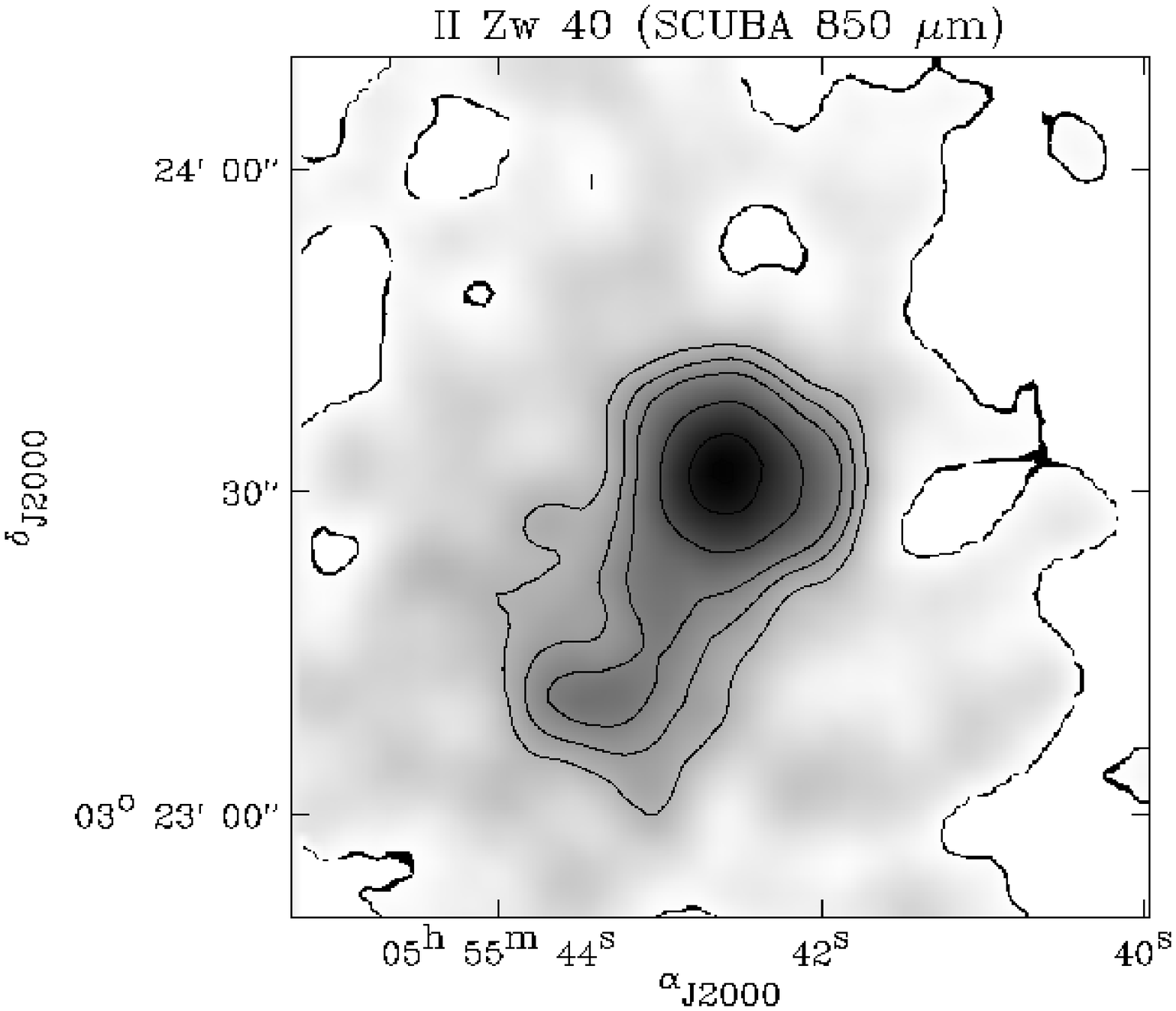}  & \hspace*{-0.6cm}
    \includegraphics[width=0.5\linewidth]{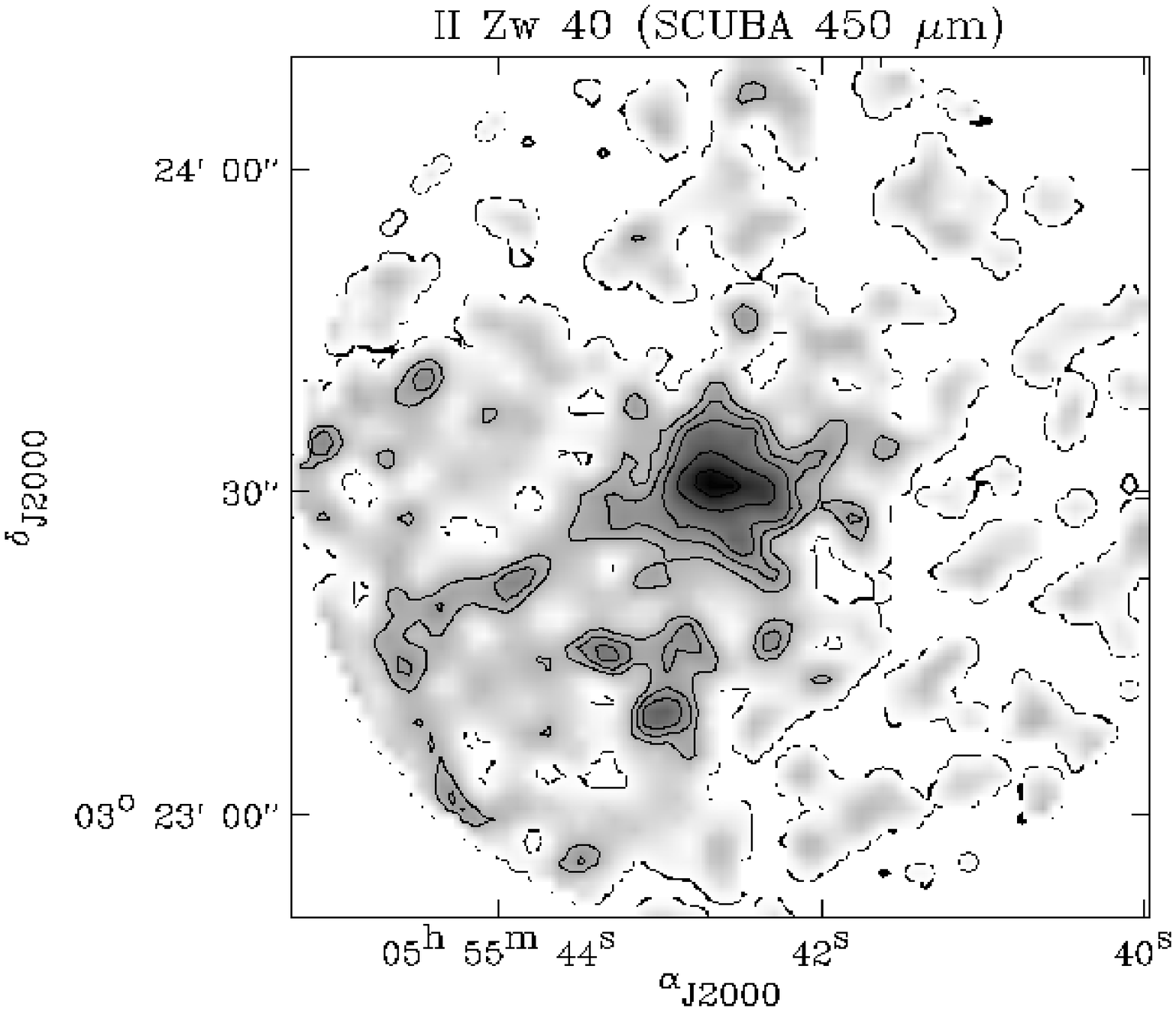}  \\
    \includegraphics[width=0.5\linewidth]{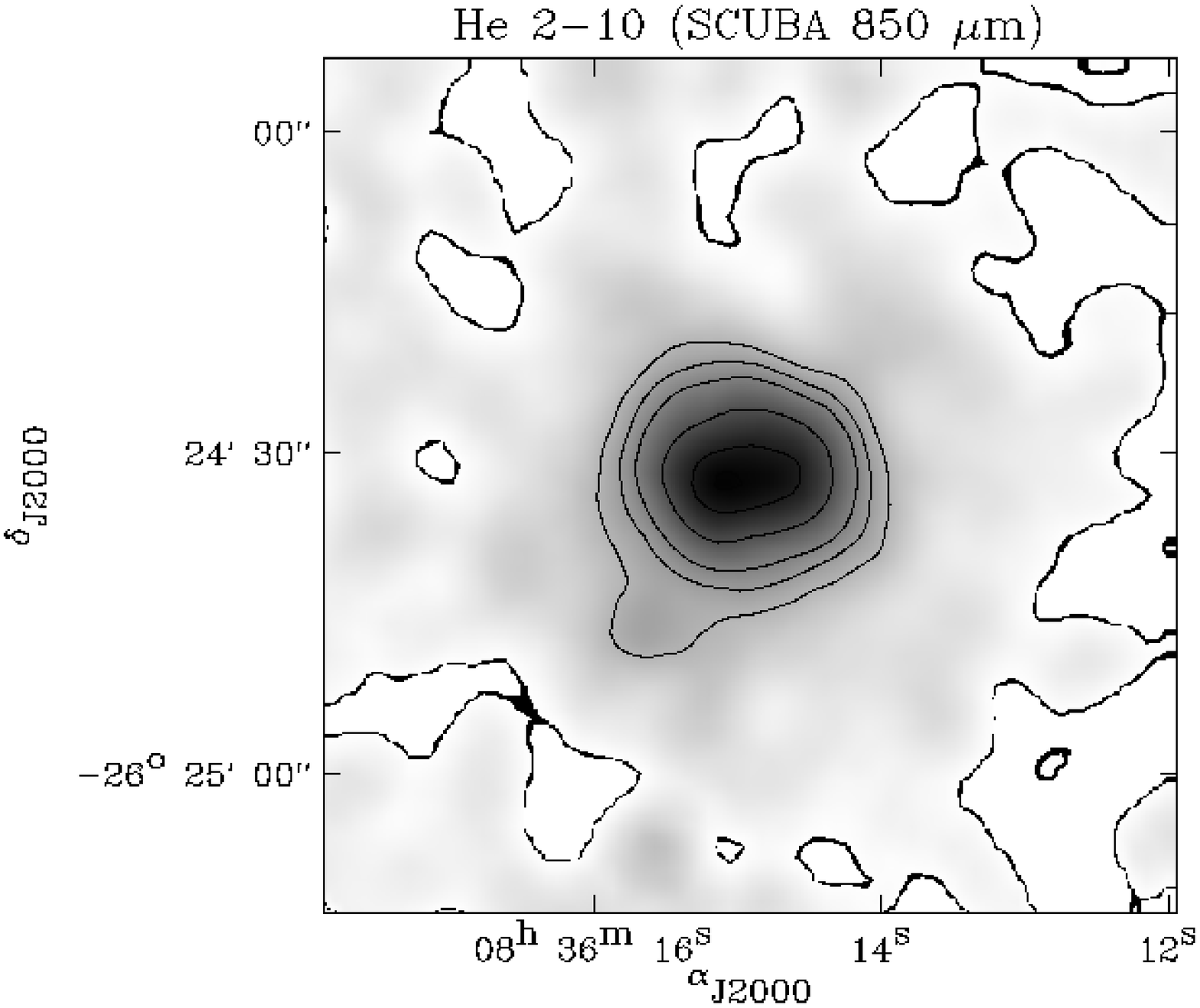}  & \hspace*{-0.6cm}
    \includegraphics[width=0.5\linewidth]{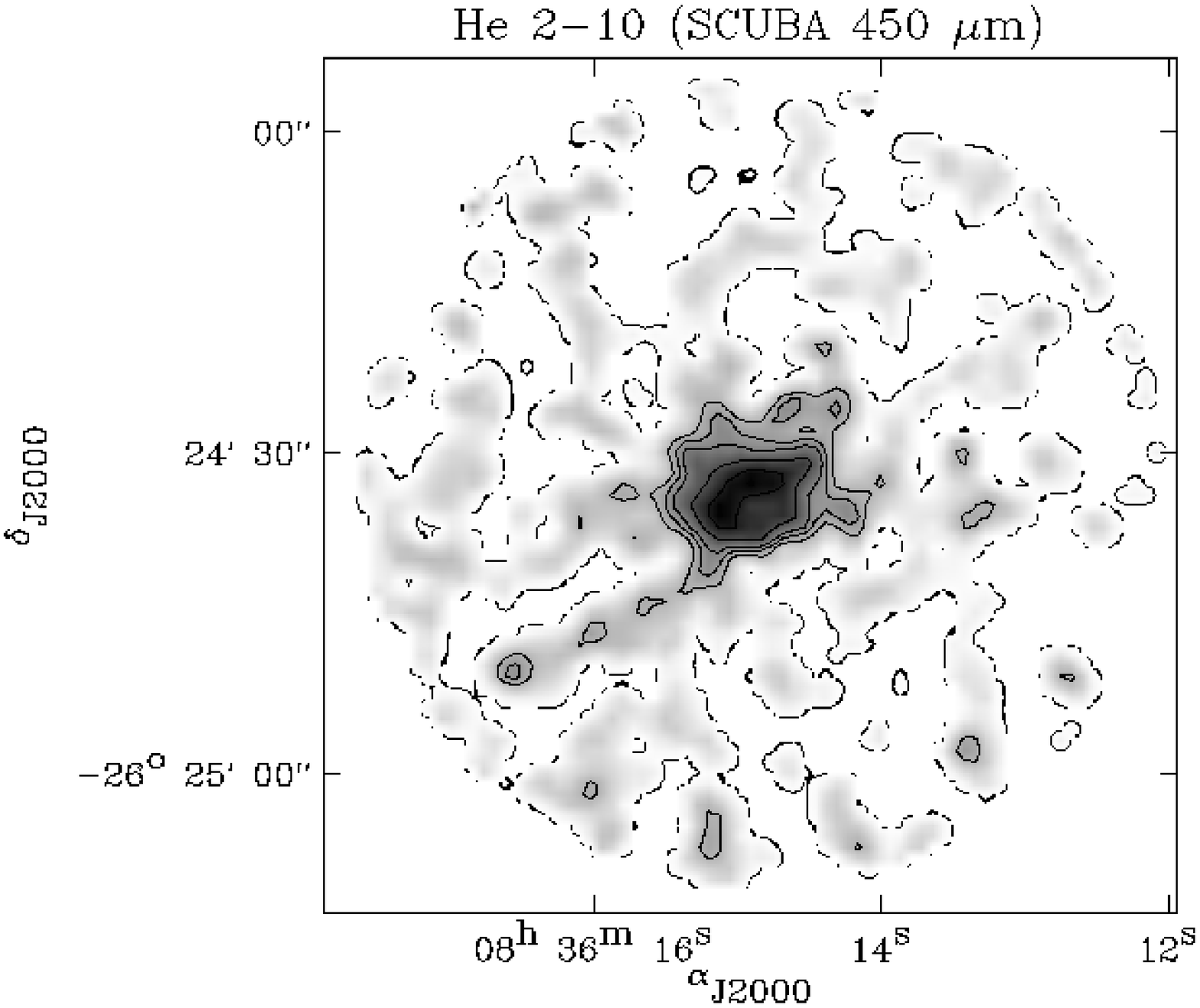}  \\
    \includegraphics[width=0.5\linewidth]{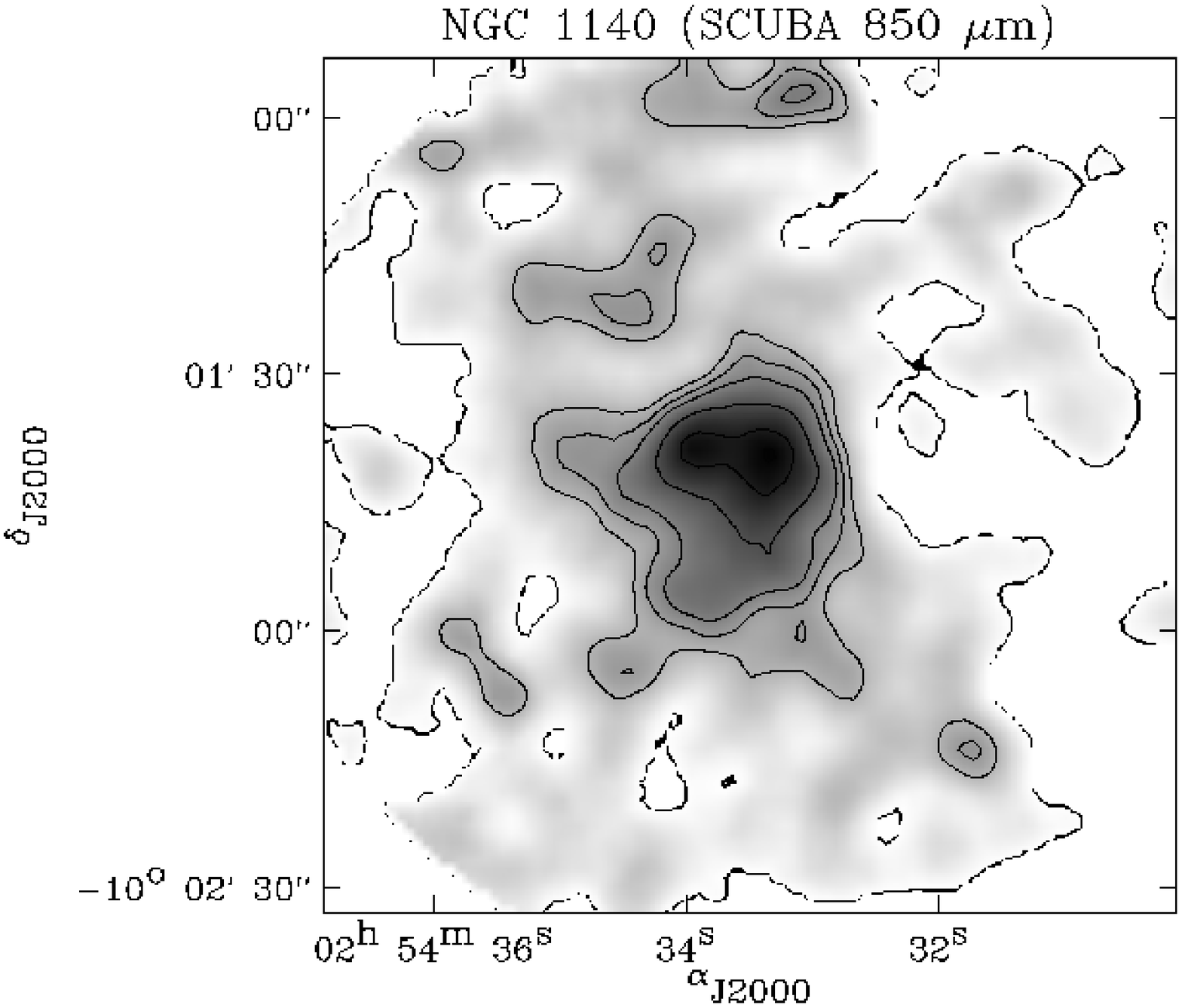} & \hspace*{-0.6cm}
    \includegraphics[width=0.5\linewidth]{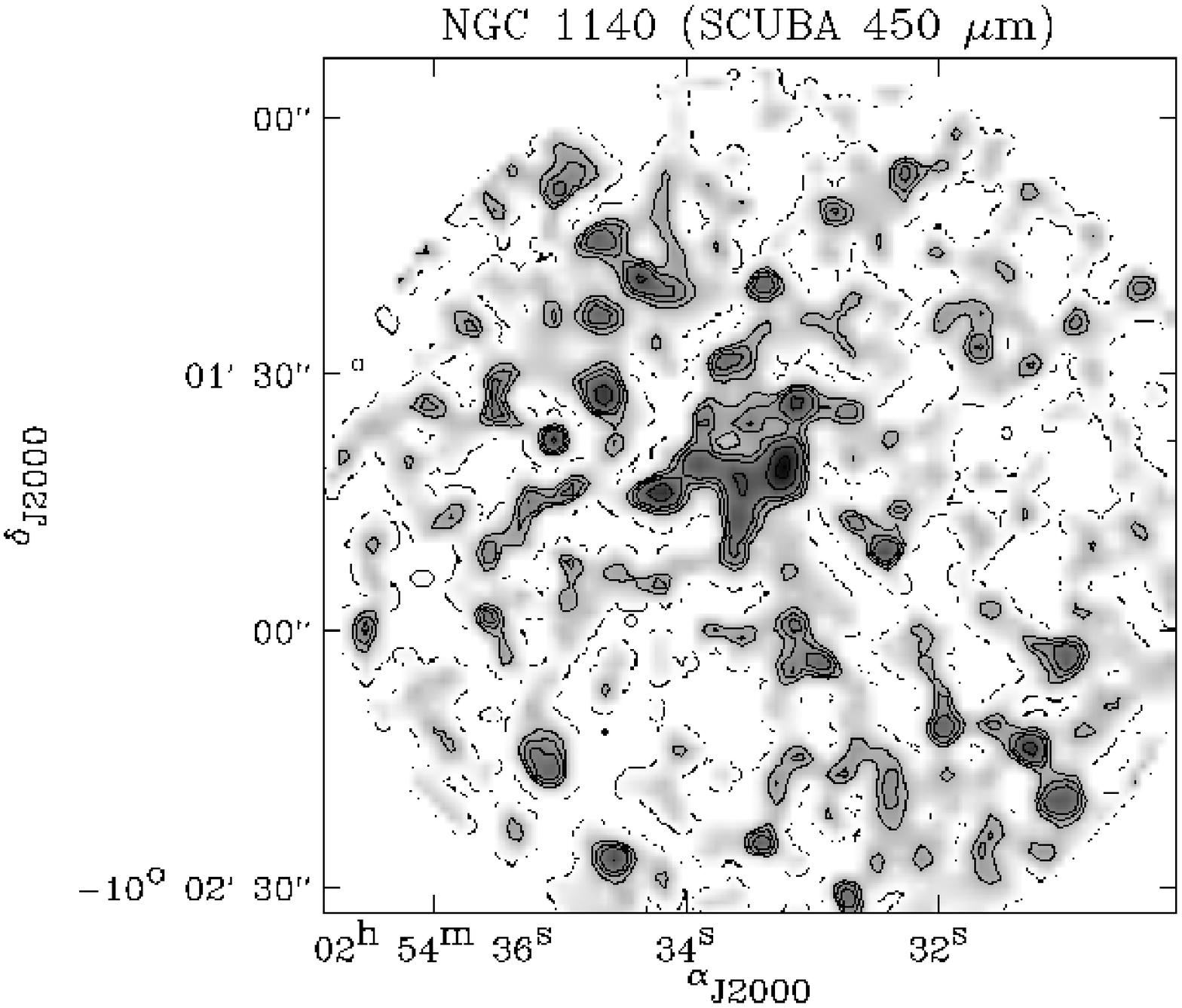} \\
  \end{tabular}
  \caption{SCUBA maps of the three galaxies.
           The first row of images is \iizw, the second is \hen, and the third
           is \ngc{1140}.
           The wavelength of the left column images is 850$\,\mic$ and the 
           wavelength of the right column images is 450$\,\mic$.
           The field of view is the same for the 450$\,\mic$ and  850$\,\mic$ 
           images of each galaxy and the color table has the same dynamic 
           range.
           The contours are 99, 90, 70, 50, 40 and 30$\,\%$ of the peak flux 
           values.}
  \label{fig:scuba}
\end{figure*}

We obtained 450 and 850$\,\mic$ data of \all, with SCUBA \citep{scuba}, a 
bolometer array on the James Clerk Maxwell Telescope (JCMT), during 
two observing runs during February 2000 and December 2000. 
Observations were carried out in the jiggle-mapping mode using a 64-point 
jiggle pattern with a chop throw of $150''$.
The precise data reduction method used is described in detail in \papii\, 
including details of the error analyses. 
The full width at half maximum (FWHM) of the beam is 8.5$''$ at $450\,\mic$ and
15.2$''$ at $850\,\mic$.
The measured $\tau_\sms{CSO}$ at 225$\,$GHz, obtained from the Caltech
Submillimeter Observatory (CSO) radiometer, ranged from 0.04 to 0.1 during our 
observations and our observed calibration sources were Uranus, Mars or 
CRL$\,$618. 
The final images are shown in Fig.~\ref{fig:scuba} and the fluxes are 
presented in Table~\ref{tab:submm}.
The signal-to-noise ratios of these images are:
$S/N(850\,\mic) \simeq 10$ and $S/N(450\,\mic) \simeq 7$, for \ngc{1140};
$S/N(850\,\mic) \simeq 15$ and $S/N(450\,\mic) \simeq 11$, for \iizw;
$S/N(850\,\mic) \simeq 23$ and $S/N(450\,\mic) \simeq 14$, for \hen.

The contamination of the 850$\,\mic$ fluxes, due to the CO(3-2) line, can be 
quantified from observations reported in the literature.
Based on CSO observations, \citet{meier+01} give an 
upper limit for \iizw\ of $I_{\rm CO(3-2)}\lesssim 0.9\rm\; K\,km\,s^{-1}$ in 
$22''$, which we convert to $F_{\rm CO(3-2)}^{850\,\mic}\lesssim 3.4\,$mJy, 
for the total galaxy by using the deconvolved 850$\,\mic$ SCUBA map and 
assuming that the CO emission has the same spatial distribution as the cold 
dust.
\citet{meier+01} also give the $I_{\rm CO(3-2)}$ for several pointings in
\hen\ which have some overlap with our SCUBA data.
We took the central measurement $I_{\rm CO(3-2)}=
16.6\rm\; K\,km\,s^{-1}$ and scaled it in the same manner as for \iizw.
The estimated CO(3-2) flux in the 850$\,\mic$ band for the total galaxy is, 
thus, $F_{\rm CO(3-2)}^{850\,\mic}=30\,$mJy (Table~\ref{tab:submm}).
In the case of \ngc{1140}, we did not find any CO(3-2) measurement, in the
literature.
However, this contribution should be very low considering the contribution of
the CO(2-1) line in the MAMBO band (Sect.~\ref{sec:mambo}), thus we believe
that this contribution is not significant.

The radio continuum contribution to our total submillimetre fluxes can
be estimated by extrapolating from the numerous radio fluxes given in
the literature.  
The radio continuum at submillimetre wavelengths is normally dominated by 
free-free emission in galaxies \citep[\eg][]{condon92}.  
The free-free emission is described by $F_\nu\propto\nu^{-0.1}$.  
In the case of \iizw, \citet{sramek+86} report a global thermal flux of 
$S_\nu=13.9-15.2\,$mJy at 4.8$\,$GHz.
We deduce a flux $S_\nu=9.1-9.9\,$mJy at 850$\,\mic$ ($\lesssim
10\,\%$) and $S_\nu=8.5-9.3\,$mJy at 450$\,\mic$ ($\lesssim 4\,\%$).
For \hen, \citet{kobulnicky+99a} report a total flux of
$F_\nu=21.1\pm1.2\,$mJy at 14.9$\,$GHz, which we convert to
$S_\nu=14.5-15.4\,$mJy at 850$\,\mic$ ($\lesssim 13\,\%$) and
$S_\nu=13.6-14.4\,$mJy ($\lesssim 4\,\%$) at 450$\,\mic$.  
We could not find a decomposition between thermal and non-thermal flux, thus
this estimate serves at the upper limit of the radio contribution in
our bands.  
We remain conservative here in order to be sure that if we do observe a 
submillimetre excess arising from dust in these galaxies,
as found previously in \ngc{1569} (\papii), the results will not be
effected by our assumptions of contamination from non-dust sources.
Finally, \citet{klein+83} report a $S_\nu=11\pm3\,$mJy flux at
10.7$\,$GHz, in \ngc{1140}.  Thus, we are able to give a range of
values of $S_\nu=5.6-9.9\,$mJy at 850$\,\mic$ ($\lesssim 12\,\%$) and
$S_\nu=4.9-8.6\,$mJy at 450$\,\mic$ ($\lesssim 4\,\%$).  These
contributions are summarised in the last part of
Table~\ref{tab:submm}.  
Thus, subtracting the radio continuum and CO contributions to the observed 
submillimetre fluxes, we are confident that we can quantify the flux density 
arising from dust only.
\begin{table}[htbp]
  \centering
  \begin{tabularx}{\linewidth}{X*{3}{l}}
  \hline
  \hline
                                   & \iizw            & \hen              & 
    \ngc{1140}          \\
                                   & (mJy)            & (mJy)             &
    (mJy)               \\
  \hline
  $F^{\rm tot}_{450\,\mic}$        & $248\pm 81$      & $342\pm 65$       & 
    $272\pm 55$          \\
  $F^{\rm tot}_{850\,\mic}$        & $98\pm 14$       & $130\pm 12$       & 
    $69\pm 28$           \\
  \hline
  CO(3-2)$_{850\,\mic}$            & $\lesssim3.4$    & $\sim 30$         & 
    -                   \\
  \hline
  Radio:                           &                  &                   &
                        \\
  - 450$\,\mic$                    & $8.5-9.3$        & $13.6-14.4$       &
    $4.9-8.6$            \\
  - 850$\,\mic$                    & $9.1-9.9$        & $14.5-15.4$       &
    $5.6-9.9$            \\
  \hline
  \end{tabularx}
  \caption{Flux and contributions to the submillimeter bands of \all.
           The total fluxes are not corrected for radio continuum and CO(3-2).}
  \label{tab:submm}
\end{table}

As noted for \ngc{1569} (\papii), the morphology of the submillimetre
emission (Fig.~\ref{fig:scuba}) is not concentrated toward the outer regions.
It follows the mid-IR emission relatively well, the peak correlating
with the star forming regions.


  \subsection{MAMBO observations}
  \label{sec:mambo}

We obtained 1.3$\,$mm data (230$\,$GHz, bandwidth$\simeq$80$\,$GHz) of \all, 
with MAMBO \citep{mambo}, a bolometer array on the IRAM 30 meter 
radio-telescope, during two observing runs in December 2001 and January 2002.
Observations were carried out in \oo\ mode using the 117-bolometer array.

\begin{figure}[htbp]
  \centering
  \includegraphics[width=\linewidth]{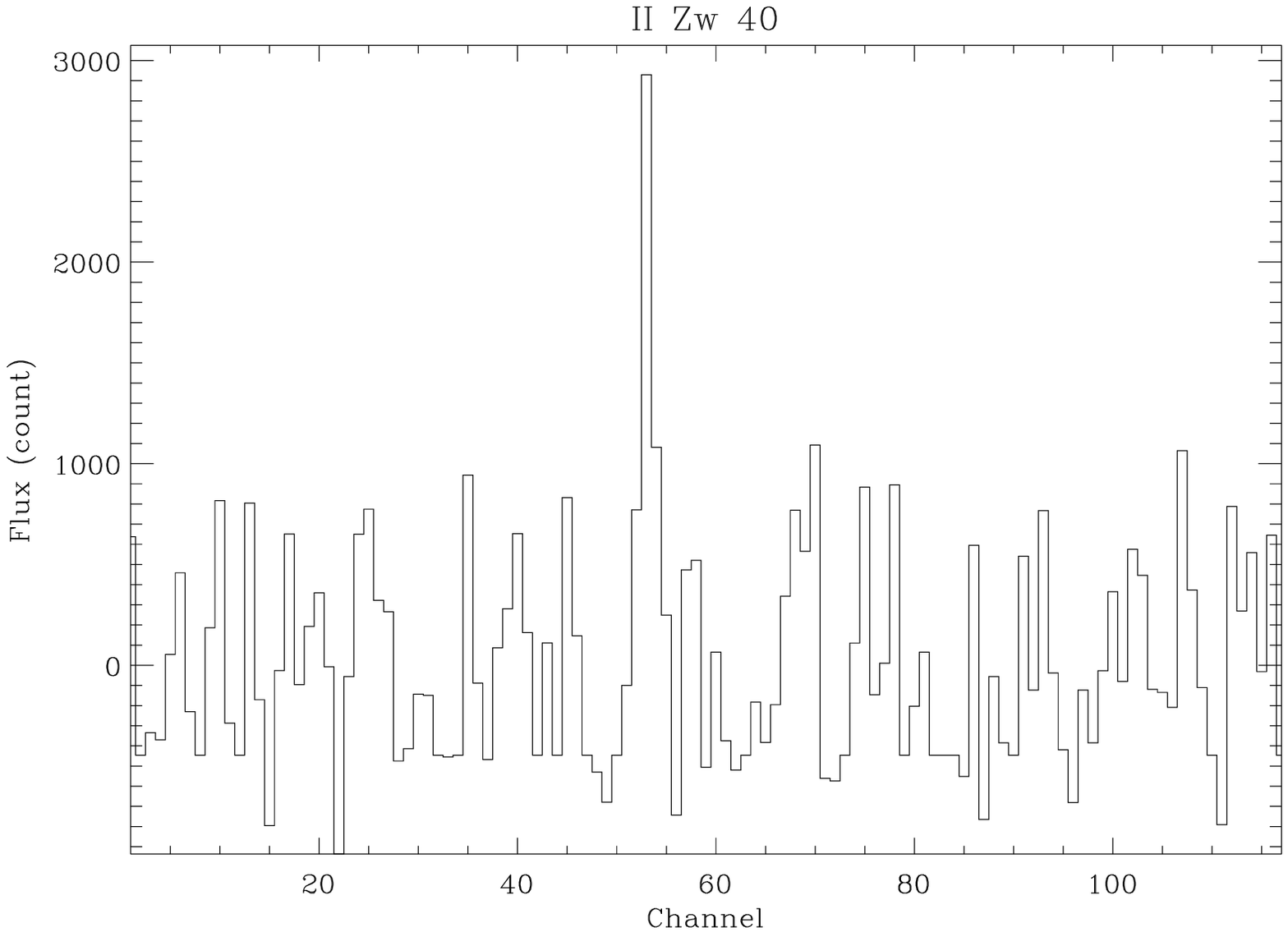} \\
  \includegraphics[width=\linewidth]{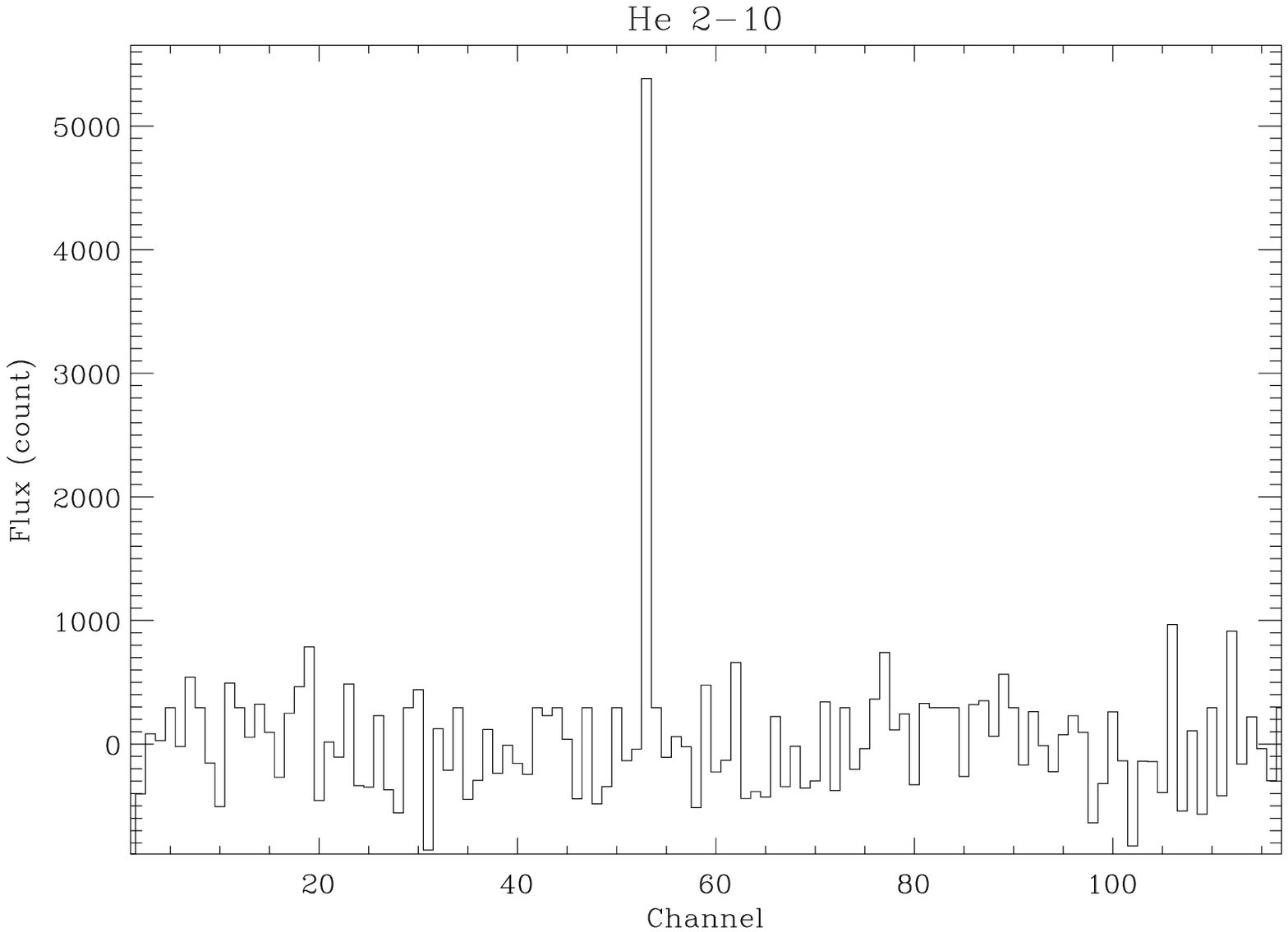}
  \caption{\oo\ observations for the two detected galaxies, \iizw\ and \hen.
           These plots show the signal in each bolometer after
           foreground subtraction.
           The peak corresponds to the reference channel (\num\ 53)
           which is centered on the source.
           The other channels are assumed to see the sky.}
  \label{fig:onf}
\end{figure}

The \oo\ observations were conducted in standard chop-nod mode, with individual
scans divided in 6 subscans each of which yielded 20 seconds of on+off source 
exposure.
The secondary mirror was chopped by 55$''$ in azimuth for \iizw\ and
\hen\ and 70$''$ for \ngc{1140}.
The pointing and the focus of the telescope were checked every hour.
The data were analysed with the NIC software \citep{nic} from the 
GILDAS package.
The data reduction included the following steps.
\\
1) Atmospheric opacity: 
The skydip measurements were carried out every two hours. 
The \tauiram, optical depth of the atmosphere at 230$\,$GHz, varied between
0.14 and 0.35 during the observations and was below 0.2 during most of the 
scans.
Moreover, it was relatively constant between successive skydips.
Each scan was corrected for atmospheric extinction using linear interpolation 
of the opacity measurements obtained from the skydip preceding the scan and 
that following the scan.
\\
2) Gain and bad bolometers:
Some bolometers are known to be noisy.
We flagged these currently-identified bad bolometers during the remainning 
procedures.
\\
3) Spikes: 
A despiking function is applied to each bolometer, with a threshold of 
$5\times\sigma$.
The spikes found are removed and replaced by the interpolation of the 
adjoining data.
\\
4) Final signal:
The mean signal of each bolometer and its variance is computed from the
corresponding subscans.
A baseline is removed from the raw signal by a least-square fit of a straight 
line to the weighted sequence of ON and OFF means.
The high-frequency noise is removed from all of the channels.
Finally, the flux of the reference channel is determined by removing the flux 
density of the adjacent four channels.
If $\mathcal{F}_{ib}$ is the flux in counts of the scan $i$ in the bolometer
$b$, we calculate the mean, 
$\mathcal{F}_b=\langle\mathcal{F}_{ib}\rangle_{\rm scans}$, of every 
scan flux, $\mathcal{F}_{ib}$, weighted by $1/\sigma_i^2$, where 
$\sigma_i=\sigma(\mathcal{F}_{ib})_{\rm bolo}$ is the deviation of each scan.
We subtract the foreground emission which is determined from the mean flux of 
the bolometers which are not on the source. 
Thus, the final signal of each bolometer is $\mathcal{F}_b^{\rm corr}=
\mathcal{F}_b-\left\langle\mathcal{F}_b\right\rangle_{b-\{\rm source\}}$.
These $\mathcal{F}_b^{\rm corr}$ are plotted in Fig.~\ref{fig:onf} for the
two galaxies which are detected.
\\
5) Calibration:
Mars, $\alpha\,$Ori, \ngc{7538} and HL$\,\tau$ were observed daily as primary 
and secondary calibrators. 
The calibration data were reduced in the same way as other galaxies.
We compute a flux conversion factor, $\Phi$, which provides the 
relationship between counts and astrophysical fluxes (in Jy), averaged over 
the calibrators.
The flux of our sources is converted to astrophysical flux values 
(in Jy/beam), 
$F_b = \Phi\times\mathcal{F}_b^{\rm corr}$, by multiplying the signal by the 
conversion factor.
We use the submillimeter spatial distribution to scale this flux to the total 
flux of each galaxies.
First, we deconvolve the SCUBA 850$\,\mic$ maps with a multi-resolution Lucy
algorithm from the MR/1 package \citep{starckbook}, using a $3\sigma$ 
detection threshold.
Second, we convolve this image to the IRAM 30-meter beam and we extract the
ratio of the total flux to the flux in one beam, 
$\mathcal{R}=F_\nu^{\rm tot}(850 \mic)/F_{\rm beam(IRAM)}(850 \mic)$.
We multiply the IRAM flux in Jy/beam by this ratio to obtain the total flux 
at 1.3$\,$mm, 
$F_\nu^{\rm tot}(1.3\,{\rm mm}) = F_{b_{\rm ref}}\times\mathcal{R}$, where 
$b_{\rm ref}$ is the reference bolometer.
The final fluxes are given in Table~\ref{tab:mm}; the values of $\mathcal{R}$
is given in the last part of this table.
\\
6) Error estimation: 
The error on the total flux is given by $\Delta F_\nu^{\rm tot}(1.3\,{\rm mm})=
\Delta F_{b_{\rm ref}}\mathcal{R} + F_{b_{\rm ref}}\Delta\mathcal{R}$.
The error on the aperture, $\Delta\mathcal{R}$, is estimated by shifting the 
SCUBA image convolved by the IRAM beam by $\pm 3''$ in RA and DEC.
This $3''$ error is the typical IRAM pointing error \citep{radiramUG} and is 
consistent with deviations observed during the pointing verification 
measurements during the observations.
The error on the flux in one beam, $\Delta F_{b_{\rm ref}}$, is the quadratic
sum of two components: $(\Delta F_{b_{\rm ref}})^2=
\sigma^2(F_{b-\{{\rm source}\}})+(\Delta\Phi/\Phi)^2$.
The first component, $\sigma(F_{b-\{{\rm source}\}})$, is the RMS of the 
final signal.
The second component, $\Delta\Phi/\Phi$, is the relative deviation of the 
flux conversion factor over the various calibrators.
The various contributions to the error are given in percentage in 
Table~\ref{tab:mm}.
\\
7) Non-dust contamination: The expected contamination from the CO(2-1)
line can be quantified from measurements in the literature.  
In the case of \iizw, we estimate 0.1$\,$mJy due to the CO(2-1) line in the
IRAM 1.3$\,$mm broadband from \citet{sage+92}.  
For \hen, we estimate a flux of 4$\,$mJy in a $12.5''$ aperture from
\citet{baas+94}, and scale it to $\sim 5\,$mJy for the total galaxy.
The upper limit given by \citet{hunter+93}, for \ngc{1140}, gives a CO(2-1) 
flux in the IRAM band of $\lesssim 0.02\,$mJy.
The radio continuum component can be extrapolated from radio fluxes that we 
used in Sect.~\ref{sec:scuba}.
In the case of \iizw, we deduce a flux $S_\nu\simeq 9.5-10.3\,$mJy at 
230$\,$GHz ($\lesssim24\,\%$); for \hen, $S_\nu\lesssim 15.1-17.0\,$mJy at 
230$\,$GHz ($\lesssim28\,\%$) and $S_\nu\simeq 5.9-10.3\,$mJy at 230$\,$GHz 
for \ngc{1140} ($\simeq20\,\%$).
These contributions are summarised in Table~\ref{tab:mm}.
\begin{table}[htbp]
  \centering
  \begin{tabularx}{\linewidth}{X*{3}{l}}
  \hline
  \hline
                                   & \iizw          & \hen            & 
    \ngc{1140}           \\
  \hline
                                   & (mJy)          & (mJy)           & 
    (mJy)                \\
  $F_\nu^{\rm tot}(1.3\,{\rm mm})$ & $43\pm 13$     & $60\pm 14$      & 
    $\lesssim 48$        \\
  Radio cont.                      & $9.5-10.3$     & $15.1-17.0$     &
    $5.9-10.3$           \\
  CO(2-1) line                     & 0.1            & 4               &
    $\lesssim 0.02$      \\
  \hline
  RMS error                        & $17\,\%$       & $2\,\%$         &
    $100\,\%$            \\
  $\Delta\Phi/\Phi$                & $16\,\%$       & $12\,\%$        &
    $16\,\%$             \\
  $\Delta\mathcal{R}$              & $22\,\%$       & $25\,\%$        &
    $30\,\%$             \\
  \hline
  $1/\mathcal{R}$                  & $41\,\%$       & $48\,\%$        &
    $22\,\%$             \\
  \hline
  \end{tabularx}
  \caption{Contributions to the 1.3 millimeter fluxes of \all.
           The total fluxes, $F_\nu^{\rm tot}(1.3\,{\rm mm})$, are not 
           corrected for radio continuum and CO(2-1).}
  \label{tab:mm}
\end{table}


  \subsection{Infrared data}
  \label{sec:ir}

The details and the data treatment of the ISOCAM CVF \mir\ observations are 
presented in \papi. 
We found that the slope of the \mir\ spectrum is a critical factor in 
constraining the dust model, particularly in the 5 to 16$\,\mic$ wavelength 
range (\papii). 
Thus, using only the IRAS 12$\,\mic$ band does not provide sufficient 
constraints on the \mir\ SED which predominantly traces the hot dust component.
From a fit of the spectrum in \iizw, we find that the internal extinction is 
$A_V \simeq 15$ (\papi).
We use the deredenned spectrum in the modeling.
We characterised the \mir\ dust continuum by choosing a few wavelength regions 
of the CVF which do not contain aromatic bands or ionic lines 
(Table~\ref{tab:obsdust}).

IRAS fluxes for \iizw\ and \ngc{1140} are given by \citet{hunter+89} and
by \citet{melisse+94}. 
\citet{thronson+86} also give the IRAS fluxes for \iizw.
The values for the four IRAS broadbands given by these authors are consistent.
Moreover, the 12$\,\mic$ IRAS flux is consistent with the ISOCAM CVF spectrum 
integrated into this band (\papi).
For our modeling purposes, we use the values of \citet{hunter+89} since 
error bars are also provided. 
These fluxes have been color-corrected.
We use the IRAS fluxes for \hen\ from \citet{sauvage+97}.

In the case of \iizw, an additional 20$\,\mic$ flux of $1.0\pm 0.2\,$Jy is 
given by \citet{roche+91}.
For \hen, we consider the flux at 11.65$\,\mic$ reported by \citet{sauvage+97},
who also give a N band flux for \hen. 
However we use the N band flux from \citet{vacca+02} instead, since there is a
large discrepancy between this flux and that cited by \citet{sauvage+97} 
\citep[see the discussion by][]{vacca+02}.
We include an unpublished M band observation of \hen\ \citep{sauvage+05} in 
our SED.


  \subsection{Optical data}
  \label{sec:opti}

To constrain the input stellar radiation field we use optical data from the
literature.
There are several papers reporting optical observations of \all.
We choose magnitudes explicitely given for the total galaxy and where the 
authors specified the Galactic foreground extinction corrections. 
Preference was given to measurements quantifying the uncertainties.
In some cases, we were obliged to scale data which were given for an aperture 
smaller than the total size of the galaxy.

In the case of \iizw, we used B, V, R and I magnitudes from \citet{cairos+01a} 
and U magnitude from the RC3 catalog \citep{devaucouleurs+91}.
The B magnitude, given by \citet{cairos+01a} is also consistent with that
provided by \citet{heisler+94}, \citet{deeg+97} and \citet{devaucouleurs+91}.
The V magnitude of \citet{cairos+01a} is consistent with that of 
\citet{devaucouleurs+91} and the R and I magnitudes are consistent with 
\citet{deeg+97}.
The J, H and K fluxes of \iizw\, reported by \citet{vanzi+96} for an aperture 
of $27''$, were scaled to obtain values for the entire galaxy. 
We computed the scaling factors from the original images \citep{vanzi+96}.

\hen\ has been observed by \citet{johansson87} in U, B, V, R, I, J, H and K
bands in several apertures.
U, B and V observations are given for the total galaxy (aperture of 61$''$),
while the other bands have been observed in, at most, a 31$''$ aperture.
The difference is relatively small (15$\,\%$).
We scaled the 31$''$ flux measurements to fit the U, B, V bands in the total 
galaxy aperture.
The error bars were not provided by the author.
Consequently, we took the deviation of the two sets of observations reported
by the author and added the error induced by the differences of apertures.
Even with these considerations, the uncertainties we obtained are relatively 
small and may be underestimated.

For \ngc{1140}, we used U, B, V, R and I magnitudes from \citet{gallagher+87} 
and J, H and K magnitudes from \citet{hunter+85}.
\citet{buat+02} also observed \ngc{1140} around 0.1$\,\mic$, in a 
$30''\times 30''$ aperture which we believe encompasses the majority of the 
emission of the galaxy at these wavelengths, which is concentrated within 
a diameter $< 3\,$kpc \citep{hunter+94b}.

These data, summarised in Table~\ref{tab:opt}, are used to constrain the 
stellar SED.
\begin{table}[htbp]
  \centering
  \begin{tabularx}{\linewidth}{X*{3}{l}}
    \hline
    \hline
                 & \iizw\        & \hen\          & \ngc{1140}    \\ 
    Band ($\mic$)& Flux (mJy)    & Flux (mJy)     & Flux (mJy)    \\ 
    \hline
    0.36 (U)     & $15.3\pm 2.3$ & $63.2\pm 2.1$  & $15.1\pm 2.3$ \\
    0.44 (B)     & $23.8\pm 1.6$ & $90.8\pm 3.0$  & $21.2\pm 3.2$ \\
    0.55 (V)     & $32.9\pm 6.3$ & $100.7\pm 5.0$ & $23.6\pm 3.5$ \\
    0.64 (R)     & $36\pm 11$    & $122.1\pm 4.0$ & $26.8\pm 4.0$ \\
    0.79 (I)     & $29.3\pm 9.7$ & $126.5\pm 4.1$ & $35.2\pm 5.3$ \\
    1.26 (J)     & $20.6\pm 4.1$ & $147.9\pm 4.8$ & $36.4\pm 5.5$ \\
    1.60 (H)     & $21.9\pm 3.3$ & $175.4\pm 5.7$ & $40.1\pm 6.0$ \\
    2.22 (K)     & $18.5\pm 2.8$ & $130.2\pm 7.5$ & $28.6\pm 4.3$ \\
    \hline
  \end{tabularx}
  \caption{Optical data from the literature (references in the text).
           These fluxes are global values, corrected for the foreground 
           Galactic extinction but not for internal extinction.}
  \label{tab:opt}
\end{table}


\section{Self-consistent modeling of the global SED}
\label{sec:model}

The modeling of the global SEDs of \all\ is done using the exact same 
procedure as described in \papii, for the case of \ngc{1569}.
We refer to this paper for the detailed description of the procedure we use 
here.
The main steps are the following.
\begin{itemize}
  \item UV-to-optical data from the literature (Table~\ref{tab:opt}), 
    corrected for the Galactic
    extinction are used to constrain the stellar radiation field.
    This radiation field is modeled with \peg\ \citep{pegase}, a stellar 
    evolutionary synthesis model.
    We consider two instantaneous bursts of star formation representing
    the young stellar population created by the recent starburst and the old
    underlying population.
    We are only interested in the shape of the ISRF here, and the age
    is a free parameter.
    Hence, the initial metallicity is not of great concern, due
    to the age-metallicity degeneracy \citep[\eg\ ][]{leborgne+04}.
    However, for a given metallicity, several age combinations fit 
    the observations.
  \item The stellar population age degeneracy is removed by constraining a 
    photoionisation model \citep[\clo; ][]{cloudy} with the \mir\ ionic line 
    ratios that we measure (\papi).
  \item This radiation field is used to heat the dust which is modeled with
    the \citet[hereafter {\dbp}]{desert+90} model.
    We fit the observed IR SED by varying the dust size distribution.
    The solution allows us to synthesize an extinction curve self-consistent
    with the emission properties.
  \item We iterate this process, correcting the UV-to-optical data for 
    internal extinction with the synthesized extinction curve, until we reach
    an agreement between the extinction and emission.
    Only the spectral dependency of the extinction curve is taken into account.
    The global internal optical depth is deduced from the energy balance 
    between stars and dust:
    \begin{equation}
      F_\star = F_\sms{UV-opt} + F_\sms{IR-mm},
    \end{equation}
    where $F_\star$ is the intrinsic flux emitted by the stars,
    $F_\sms{UV-opt}$, the escaping stellar flux, and
    $F_\sms{IR-mm}$, the flux reemitted by the dust (these fluxes are 
    integrated over frequency).
\end{itemize}


  \subsection{The observed SEDs}
  \label{sec:obssed}

The stellar SEDs are constrained by the fluxes in Table~\ref{tab:opt} and by
\mir\ ionic line ratios (\papi).
The [\neii] (12.81$\,\mic$) line is not detected in \iizw, thus, in this case,
we constrain the photoionisation model with only two ratios: [\siv]/[\neiii] 
and [\ariii]/[\neiii].	
In the case of \hen, we do not have an ISOCAM CVF spectrum. 
However \citet{beck+97} report NASA IRTF \mir\ spectroscopy of this galaxy 
and they measure ionic line intensities.
We deduce from this study, the ratios 
$[\mbox{\neii}]/[\mbox{\ariii}] = 12.5\pm 5.2$ and
$[\mbox{\neii}]/[\mbox{\siv}] = 50\pm 40$.

The dust SEDs are constrained by the data points shown in 
Table~\ref{tab:obsdust}.
All the non-dust contributions have already been subtracted from these fluxes 
listed in Table~\ref{tab:obsdust}.
\begin{table*}[htbp]
  \centering
  \begin{tabularx}{\linewidth}{X*{3}{|XX}}
    \hline
    \hline
    & \multicolumn{2}{c}{\iizw} 
    & \multicolumn{2}{|c|}{\hen}
    & \multicolumn{2}{c}{\ngc{1140}} \\
    Instrument &
    $\lambda\;(\mic)$      & Flux (mJy)       &
    $\lambda\;(\mic)$      & Flux (mJy)       &
    $\lambda\;(\mic)$      & Flux (mJy)       \\
    \hline
    ISOCAM &
    8.8                    & $89\pm 55$       &
                           &                  &
    6.1                    & $41\pm 26$       \\
    &
    9.4                    & $132\pm 47$      &
                           &                  &
    6.2                    & $89\pm 22$       \\
    &
    10.1                   & $177\pm 64$      &
                           &                  &
    6.5                    & $46\pm 25$       \\
    &
    10.8                   & $238\pm 64$      &
                           &                  &
    6.9                    & $55\pm 21$       \\
    &
    11.8                   & $315\pm 65$      &
                           &                  &
    7.4                    & $82\pm 21$       \\
    &
    12.1                   & $339\pm 67$      &
                           &                  &
    7.6                    & $119\pm 20$      \\
    &
    13.2                   & $417\pm 73$      &
                           &                  &
    7.8                    & $113\pm 20$      \\
    &
    13.9                   & $458\pm 75$      &
                           &                  &
    7.9                    & $120\pm 21$      \\
    &
    14.6                   & $489\pm 69$      &
                           &                  &
    8.1                    & $65\pm 20$       \\
    &
    15.0                   & $504\pm 78$      &
                           &                  &
    10.9                   & $82\pm 15$       \\
    &
    16.0                   & $532\pm 87$      &
                           &                  &
    11.1                   & $149\pm 17$      \\
    &
                           &                  &
                           &                  &
    11.3                   & $192\pm 17$      \\   
    &
                           &                  &
                           &                  &
    11.5                   & $118\pm 18$      \\   
    &
                           &                  &
                           &                  &
    11.7                   & $90\pm 16$       \\   
    &
                           &                  &
                           &                  &
    13.4                   & $91\pm 23$       \\   
    &
                           &                  &
                           &                  &
    13.9                   & $62\pm 20$       \\   
    &
                           &                  &
                           &                  &
    14.4                   & $89\pm 23$       \\
    &
                           &                  &
                           &                  &
    16.1                   & $78\pm 35$       \\
    \hline
    Others &
    20 (Q)                 & $1000\pm 200$    &
    4.8 (M)                & $60\pm 9$        &
                           &                  \\
    &
                           &                  &
    10.1 (N)               & $720\pm 95$      &
                           &                  \\
    &
                           &                  &
    11.65                  & $850\pm 100$     &
                           &                  \\
    \hline
    IRAS &
                           &                  &
    12                     & $1100\pm 160$    &
                           &                  \\
    &
    25                     & $2170\pm 300$    &
    25                     & $6550\pm 980$    &
    25                     & $390\pm 70$      \\
    &
    60                     & $7280\pm 900$    &
    60                     & $23800\pm 3600$  &
    60                     & $3940\pm 600$    \\
    &
    100                    & $6360\pm 1100$   &                      
    100                    & $31800\pm 4800$  &
    100                    & $5000\pm 800$    \\
    \hline
    SCUBA &
    443                    & $241\pm 81$      &
    443                    & $328\pm 67$      &
    443                    & $270\pm 110$     \\
    &
    863                    & $87\pm 14$       &
    863                    & $85\pm 36$       &
    863                    & $69\pm 28$       \\
    \hline
    MAMBO &
    1200                   & $33\pm 13$       &
    1200                   & $40\pm 18$       &
    1200                   & $\lesssim 42$    \\
    \hline
  \end{tabularx}
  \caption{Observed dust SED.
           The data are the flux values used to constrain the dust
           model.
           They are corrected for all non-dust effects (molecular lines,
           radio continuum) and extinction.
           The references are given in the text in Sect.~\ref{sec:obs}.
           For \ngc{1140}, the ISOCAM data which are reported are those used to
           constrain both the continuum and the PAH bands.
           The VSGs have been constrained separately (see explanation in 
           text).}
  \label{tab:obsdust}
\end{table*}


  \subsection{The dust model}
  \label{sec:dustmod}

As in \papii, we define an effective radius of the galaxies, $R_{\rm eff}$, 
which is the effective distance between stars and dust in a thin
shell model.
We estimate the equivalent radius, $R_{\rm equiv}$, by fitting an ellipse
to the 850$\,\mic$ images and $R_{\rm eff} = 3/4\times R_{\rm equiv}$.

For \iizw, we assume a distance of $D=10\pm 1\,$Mpc and for \hen, we assume
$D=9\pm 1\,$Mpc, consistent with all the recent values adopted in 
the literature \citep{roche+91,conti91,vacca+92,telesco+93}.
In the case of \ngc{1140}, we adopt $D=23\pm 2\,$Mpc to be consistent with the
recent values of 20 to 25$\,$Mpc 
\citep{kinney+93,hunter+94a,saikia+94,heckman+98,buat+02}. 
The relevant model parameters for the sources are given in 
Table~\ref{tab:physpar}.
\begin{table}[htbp]
  \centering
  \begin{tabularx}{\linewidth}{X*{5}{l}}
    \hline
    \hline
    Name    & $R_{\rm eff}$  & $D$          & 
      $M(\mbox{\hi})$           & $Z/Z_\odot$       \\
            & (kpc)          & (Mpc)        & 
      ($\msol$)                 &                   \\
    \hline
    \iizw\  & $0.5\pm 0.2$   & $10\pm 1$    & 
      $4.4\times 10^{8\;a}$     & $\sim 1/6^{\;b}$  \\
    \hen\   & $0.8\pm 0.3$   & $9\pm 1$     & 
      $3.1\times 10^{8\;c}$     & $\sim 1/1^{\;d}$  \\
    \ngc{1140} & $2.6\pm 0.9$   & $23\pm 2$    & 
      $7.5\times 10^{9\;e}$     & $\sim 1/3^{\;f}$  \\
    \hline
  \end{tabularx}
  \caption{Physical parameters of \all.
           $R_{\rm eff}$ is deduced from our \mir\ and submillimetre images, 
           the other parameters are found in the literature.
           $D$ is the distance to the galaxy and $Z$, the metal abundances.
           ($a$)~\citet{vanzee+98};
           ($b$)~e.g. \citet{walsh+93}, \citet{masegosa+94};
           ($c$)~\citet{sauvage+97};
           ($d$)~\citet{kobulnicky+99b};
           ($e$)~\citet{hunter+94a};
           ($f$)~\citet{guseva+00}.
           }
  \label{tab:physpar}
\end{table}

The \dbp\ dust model was originally designed to explain the dust properties of
the Milky Way and includes three dust components:
the Polycyclic Aromatic Hydrocarbons (PAHs) which are responsible for the 
\mir\ emission features at 3.3, 6.2, 7.7, 8.6 and 11.3$\,\mic$,
the Very Small Grains (VSGs), which are assumed to be the carriers of the 
2175$\,$\AA\ extinction bump and the Big Grains (BGs), which have an emission 
peak in the \fir. 
VSGs are carbonaceous grains and BGs are silicate grains.
Each of the 3 dust components is described by their minimum and maximum grain 
size (a$_{-}$ and a$_{+}$), and $\alpha$, the index of the power law size 
distribution which is $n(a)\propto a^{-\alpha}$ ($a$ is the grain radius and 
$n(a)$ is the number density of grains between $a$ and $a+da$). 
The dust mass abundance of each component is $Y=M_\sms{X}/M_\sms{H}$, where 
$M_\sms{X}$ is the mass of the dust component, $X$, and $M_\sms{H}$ is the 
hydrogen mass in the galaxy.


\section{Results and discussion}
\label{sec:results}


  \subsection{The model parameters}
  \label{sec:param}

\begin{figure}[htbp]
  \begin{center}
  \includegraphics[width=\linewidth]{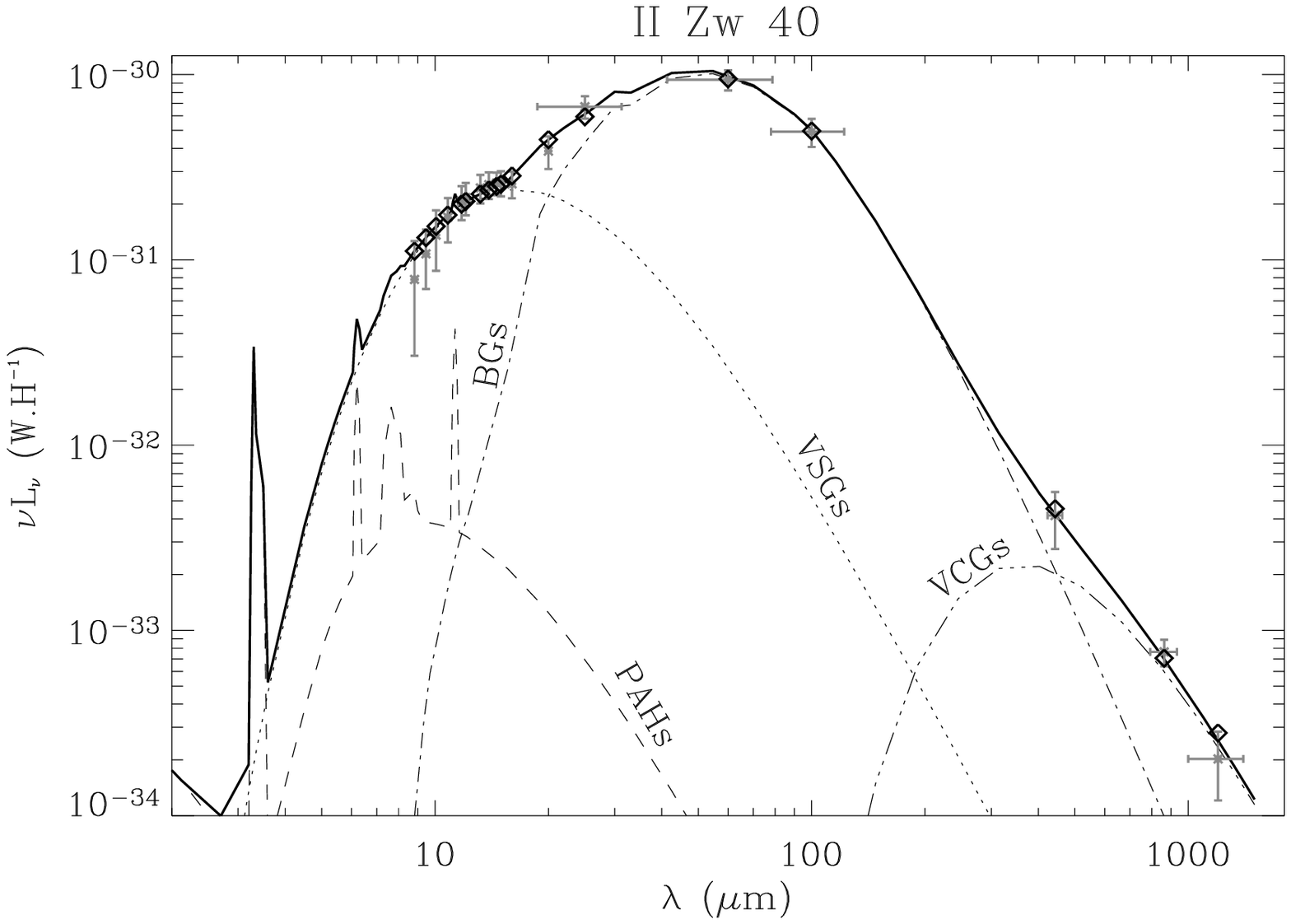} \\
  \includegraphics[width=\linewidth]{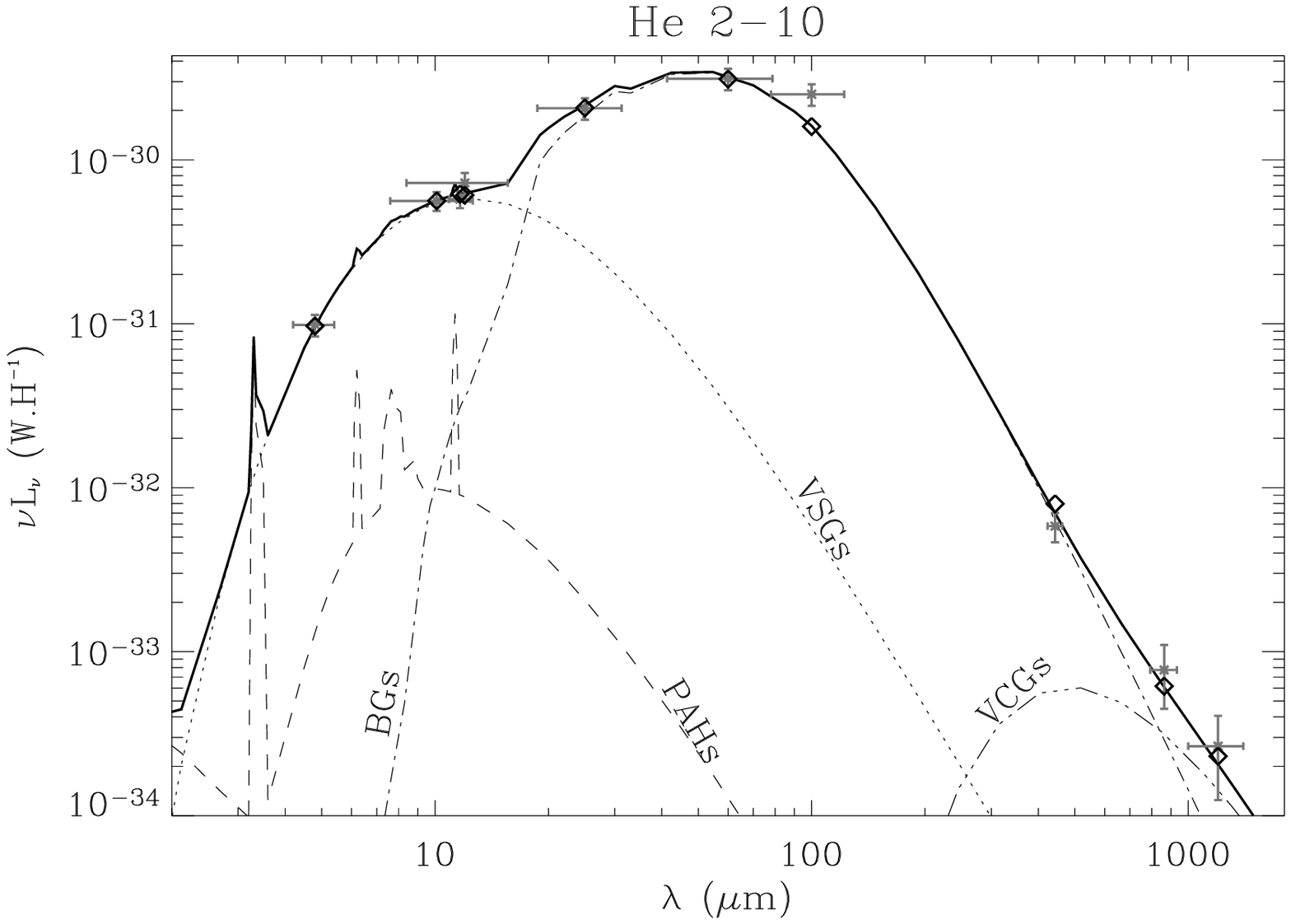} \\
  \includegraphics[width=\linewidth]{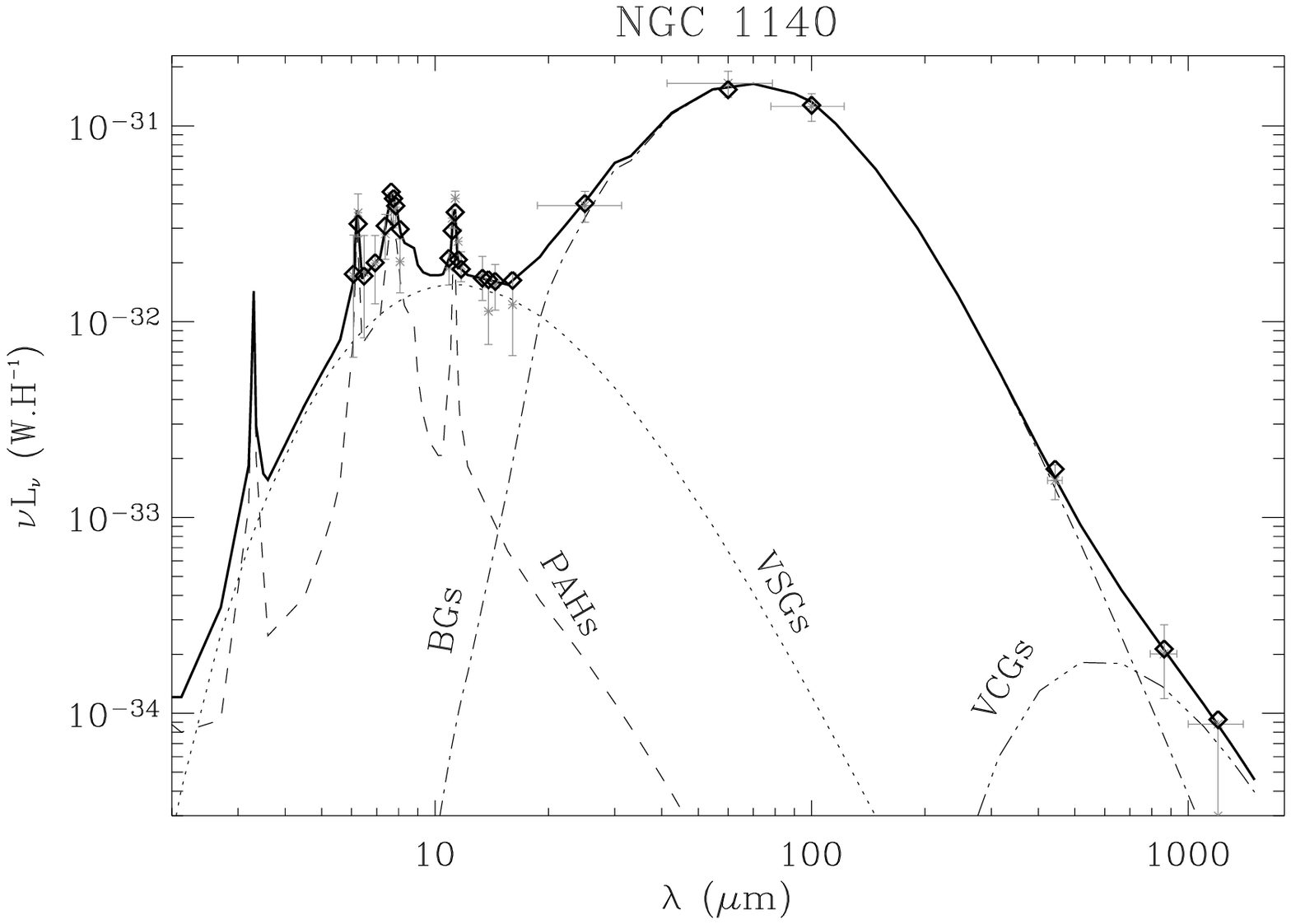}
  \caption{\all\ observations and modeled SED.
           The data (Table~\ref{tab:obsdust}) are indicated by crosses: 
           vertical bars are the errors on the flux values and the horizontal 
           bars indicate the widths of the broadbands.
           The lines are the dust model and the different dust components.
           Diamonds are the model integrated over the observational broadbands 
           and color-corrected.
           Thus, departures from the model lie where the diamonds deviate 
           from the crosses.
           The power is expressed in $\rm W\, H^{-1}$ which is $\rm \nu L_\nu$
           divided by the number of H atoms ($\rm L_\nu$ is the monochromatic
           luminosity).}
  \label{fig:sed}
  \end{center}
\end{figure}

We attempt to fit the observed SED with the \dbp\ dust model, varying the
dust parameters for each component.
We retain the original denomination of BGs, even if the model results indicate
that these grains are small, as in the cases reported here. 
The values of the parameters of the best $\chi^2$ fits to the SEDs
(Fig.~\ref{fig:sed}) are given in Table~\ref{tab:paramod}.
\begin{table*}[htbp]
  \centering
  \begin{tabularx}{\linewidth}{*{7}{X}}
    \hline
    \hline
            &          & \mw\                  & \ngc{1569}                   &
      \iizw\                     & \hen\                
      & \ngc{1140}          \\
    \hline
    PAHs    & $Y$      & $4.3\times 10^{-4}$   & $\lesssim 1\times 10^{-6}$   &
      $\lesssim 5\times 10^{-7}$ & $\lesssim 3\times 10^{-6}$
      & $8.3\times 10^{-7}$ \\
            & $a_-$    & $0.4\,$nm             & $0.4\,$nm                    &
      $0.4\,$nm                  & $0.4\,$nm               
      & $0.4\,$nm           \\
            & $a_+$    & $1.2\,$nm             & $1.2\,$nm                    &
      $1.2\,$nm                  & $1.2\,$nm              
      & $1.2\,$nm           \\
            & $\alpha$ & 3.0                   & 3.0                          &
      3.0                        & 3.0                  
      & 3.0                 \\
    \hline
    VSGs    & $Y$      & $4.7\times 10^{-4}$   & $1.8\times 10^{-5}$          &
      $2.3\times 10^{-5}$        & $4.9\times 10^{-5}$             
      & $2.2\times 10^{-5}$ \\
            & $a_-$    & $1.2\,$nm             & $1.2\,$nm                    &
      $2.0\,$nm                  & $1.2\,$nm                
      & $1.2\,$nm           \\
            & $a_+$    & $15\,$nm              & $7.8\,$nm                    &
      $2.9\,$nm                  & $2.6\,$nm                
      & $3.0\,$nm           \\
            & $\alpha$ & 2.6                   & 4.0                          &
      1.0                        & 5.5                     
      & 7.9                 \\
    \hline
    BGs     & $Y$      & $6.4\times 10^{-3}$   & $4.4\times 10^{-4}$          &
      $2.6\times 10^{-4}$        & $8.9\times 10^{-4}$             
      & $6.1\times 10^{-4}$ \\
            & $a_-$    & $15\,$nm              & $2.2\,$nm                    &
      $3.2\,$nm                  & $2.5\,$nm                
      & $3.8\,$nm           \\
            & $a_+$    & $110\,$nm             & $110\,$nm                    &
      $110\,$nm                  & $110\,$nm                
      & $110\,$nm           \\
            & $\alpha$ & 2.9                   & 6.3                          &
      25                         & 8.9                     
      & 30                  \\
    \hline
    VCGs    & $Y$      & \it unknown           & $1.3\times 10^{-3}$          &
      $1.3\times 10^{-3}$        & $1.5\times 10^{-3}$             
      & $\sim 1.1\times 10^{-3}$\\
            & $T$      & \it unknown           & 5$\,$K                       &
      8$\,$K                     & 6$\,$K                     
      & $\sim 5\,$K         \\
            & $\beta$  & \it unknown           & 1                            &
      1                          & 1                     
      & $\sim 1$            \\
    \hline
     Statistics       & $n$        &  -  &   18 &   18 &   10 &   24 \\
                      & $m$        &  -  &    7 &    7 &    6 &    7 \\
                  & $m_\sms{VCG}$  &  -  &    9 &    9 &    8 &   10 \\
           & $\bar{\chi^2}$        &  -  &  1.7 &  4.5 &  9.8 &  1.5 \\
       & $\bar{\chi^2}_\sms{VCG}$  &  -  & 0.27 & 0.36 &  2.8 & 0.89 \\
                 & $\Delta AIC$    &  -  &   46 &   67 &   25 &   52 \\
                 & $\Delta AICC$   &  -  &   61 &   82 &  149 &   58 \\
    \hline
  \end{tabularx}
  \caption{Dust parameter values and statistical quantities, for the best 
           model fit.
           For comparison, we give the corresponding values for the \mw\ 
           (\dbp) and for \ngc{1569} (\papii).}
  \label{tab:paramod}
\end{table*}

The bulk of the emission originates essentially in the VSGs and BGs 
(Fig.~\ref{fig:sed}).
The free parameters are $a_+$, $\alpha$ and $Y$, for the VSGs, and
$a_-$, $\alpha$ and $Y$, for the BGs.
We fix $a_+^\sms{BG}$ to the Galactic value, since its determination from
the fit is very uncertain, the large sizes not being significant contributers 
to the emission.
For \hen\ and \ngc{1140}, we also fix $a_-^\sms{VSG}$ to the Galactic value,
since the best $\chi^2$ is obtained for very low unphysical values (a few \AA).
On the other hand, for \iizw, the value of $a_-$ can be determined.
Among our sources, \hen\ is the most poorly constrained, we do not have
a CVF spectrum.
However the number of free parameters remains smaller than
the number of observations (Table~\ref{tab:paramod}).
Since the PAH emission is detected only in \ngc{1140}, $Y_\sms{PAH}$ remains
a free parameter.
For \iizw\ and \hen, we are able to put only an upper limit on the PAH mass.

Our results show that the three component DBP90 dust model is not adequate to 
explain the 
full dust SED out to millimetre wavelengths for any of the galaxies, as we 
first discovered in \ngc{1569} (see \papii\ for this point).
We are compelled to add a fourth component for each 
of the galaxies. 
The Very Cold Grains (VCGs) are modeled using a modified black 
body, with emissivity indices of $\beta = 1$ (consistent 
with carbonaceous grain properties) and $\beta = 2$ (consistent with silicate 
grain properties). 
This very cold dust component is constrained, only by the three submillimetre
observations.
Thus it is very difficult to constrain the temperature, the mass and the 
$\beta$ of this component.
We choose to vary only the temperature $T$ and the mass of this component,
considering the different values of $\beta$.
However, for \iizw\ and \hen, the fit with $\beta = 2$ gives a dust mass
higher than the metal mass of the galaxy, thus we exclude this value and fit
only a $\beta = 1$ black body. 
In the case of \ngc{1140}, we only have an upper limit on the flux at 
1.3$\,$mm, so we do not have sufficient data to constrain the value of $\beta$.

Table~\ref{tab:paramod} contains the values of several statistical quantities:
$n$ is the number of observations, $m$, the number of free parameters used for
the standard model, $m_\sms{VCG}$, the number of free parameters after the
introduction of the VCGs, $\bar{\chi^2} = \chi^2/(n-m-1)$ is the reduced chi 
square, for the standard model, and $\bar{\chi^2}_\sms{VCG}$, the reduced chi 
square with VCGs.

The introduction of the VCGs, having important astrophysical consequences,
should be examined with scrutinity.
Hence, we have calculated the Akaike's Information Criterion (AIC), in order to
examine the validity of introducing this new component.
The AIC is a statistical quantity that can be regarded as the extension of the
log likelihood, including the effect of the instability induced by the number 
of parameters.
An elementary derivation of the AIC can be found in \citet{takeuchi00}, and
a brief summary of its use in Appendix of \citet{takeuchi+00}.
The squared sum of the deviations of the model from the data is:
\begin{equation}
  S = \sum_{i=1}^{n}[\log(L_{\nu i})-\log(f(\lambda_i,\theta))]^2,
\end{equation}
$L_{\nu i}$ being the observed monochromatic luminosity at the wavelength
$\lambda_i$, $f(\lambda_i,\theta)$, the value of the model at the same 
wavelength, and $\theta$, the parameter vector.
The AIC is defined by:
\begin{equation}
  AIC = n\times\ln(2\pi+1) + n\times\ln\left(\frac{S}{n}\right) + 2(m+1).
\end{equation}
The second order corrected AICC is:
\begin{equation}
  AICC = AIC + \frac{2(m+1)(m+2)}{n-m-2}.
\end{equation}
A variation of AIC and AICC greater than unity, after introducing the VCGs, 
is required to justify this new component.
Table~\ref{tab:paramod} contains the values of 
$\Delta AIC = AIC_\sms{standard} - AIC_\sms{VCG}$ and $\Delta AICC$ which is
defined identically.
We see that these variations are much larger than unity, thus the VCG component
is statistically reasonable.

The range of reliability of the parameters is provided in 
Table~\ref{tab:paramerr}.
The uncertainties of the parameters in Table~\ref{tab:paramerr} are derived 
from both the uncertainty of the geometry (the value of $R_{\rm eff}$) and the 
errors of the observations (Section~\ref{sec:obs}).
A complete description of the way we compute these errors is given in \papii\ 
(Sect.~4.1.2).
\begin{table*}[htbp]
  \centering
  \begin{tabularx}{\linewidth}{*{6}{X}}
    \hline
    \hline
            &          & \ngc{1569}                         &
      \iizw\                 & \hen\                  &
      \ngc{1140}             \\
    \hline
    PAHs    & $Y$      & $\sim 0-10^{-6}$             &
      $\sim 0-5\times 10^{-7}$  & $\sim 0-3\times 10^{-6}$  & 
      $(0.5-1.4)\times 10^{-6}$   \\
    \hline
    VSGs    & $Y$      & $(1.3-2.3)\times 10^{-5}$    &
      $(1.4-3.8)\times 10^{-5}$ & $(2.8-7.5)\times 10^{-5}$ & 
      $(1.1-2.8)\times 10^{-5}$   \\
            & $a_-$    & $1.2\,$nm \it fixed               &
      $1.1-2.5\,$nm             & $1.2\,$nm \it fixed           & 
      $1.2\,$nm \it fixed              \\
            & $a_+$    & $3.5-12\,$nm                 &
      $2.1-5.0\,$nm             & $1.9-2.7\,$nm             & 
      $2.0-3.0\,$nm               \\
            & $\alpha$ & $2.6-5.2$                    &
      $0.2-3.7$                 & $2.8-7.1$                 & 
      $3.2-36$                    \\
    \hline
    BGs     & $Y$      & $(3.5-4.7)\times 10^{-4}$    &
      $(1.9-4.0)\times 10^{-4}$ & $(0.5-1.5)\times 10^{-3}$ & 
      $(0.3-1.1)\times 10^{-3}$   \\
            & $a_-$    & $2.1-2.9\,$nm                &
      $3.0-3.7\,$nm             & $1.7-2.6\,$nm             &  
      $3.7-4.0\,$nm               \\
            & $a_+$    & $110\,$nm \it fixed                       &
      $110\,$nm \it fixed                 & $110\,$nm \it fixed                 &  
      $110\,$nm \it fixed                   \\
            & $\alpha$ & $5.8-35$                     &
      $25-37$                   & $4.8-15.5$                & 
      $23-40$                     \\
    \hline
    VCGs    & $Y$      & $(1.3-0.4)\times 10^{-3}$    &
      $(2.0-0.7)\times 10^{-3}$ & $(0.7-4.1)\times 10^{-3}$ & 
      $(1.7-0.2)\times 10^{-3}$   \\
            & $T$      & $5-7$ K                      &
      $7-9$ K                   & $5-7$ K                   &
      $5-9$ K                     \\
            & $\beta$  & 1  \it fixed                           &
      1     \it fixed                     & 1    \it fixed                      &
      $1-2$                       \\
    \hline
  \end{tabularx}
  \caption{Range of reliability of the dust parameters.
           For comparison, we give the corresponding values for \ngc{1569} 
           (\papii).
           These limits take into account both the spread due to the error 
           bars of the observations and the spread due to the assumed value 
           of the radius.
           The indices of the power-law distribution, $\alpha$, have high 
           uncertainties since the size distribution is very steep.
           For these cases, the choice of a single grain size would 
           effectively fit the emission.
           ``Fixed'' indicates that these parameters are fixed to the Galactic
           values due to the lack of constraints.}
  \label{tab:paramerr}
\end{table*}


  \subsection{The size distributions}
  \label{sec:dist}

\begin{figure}[htbp]
  \begin{center}
  \includegraphics[width=\linewidth]{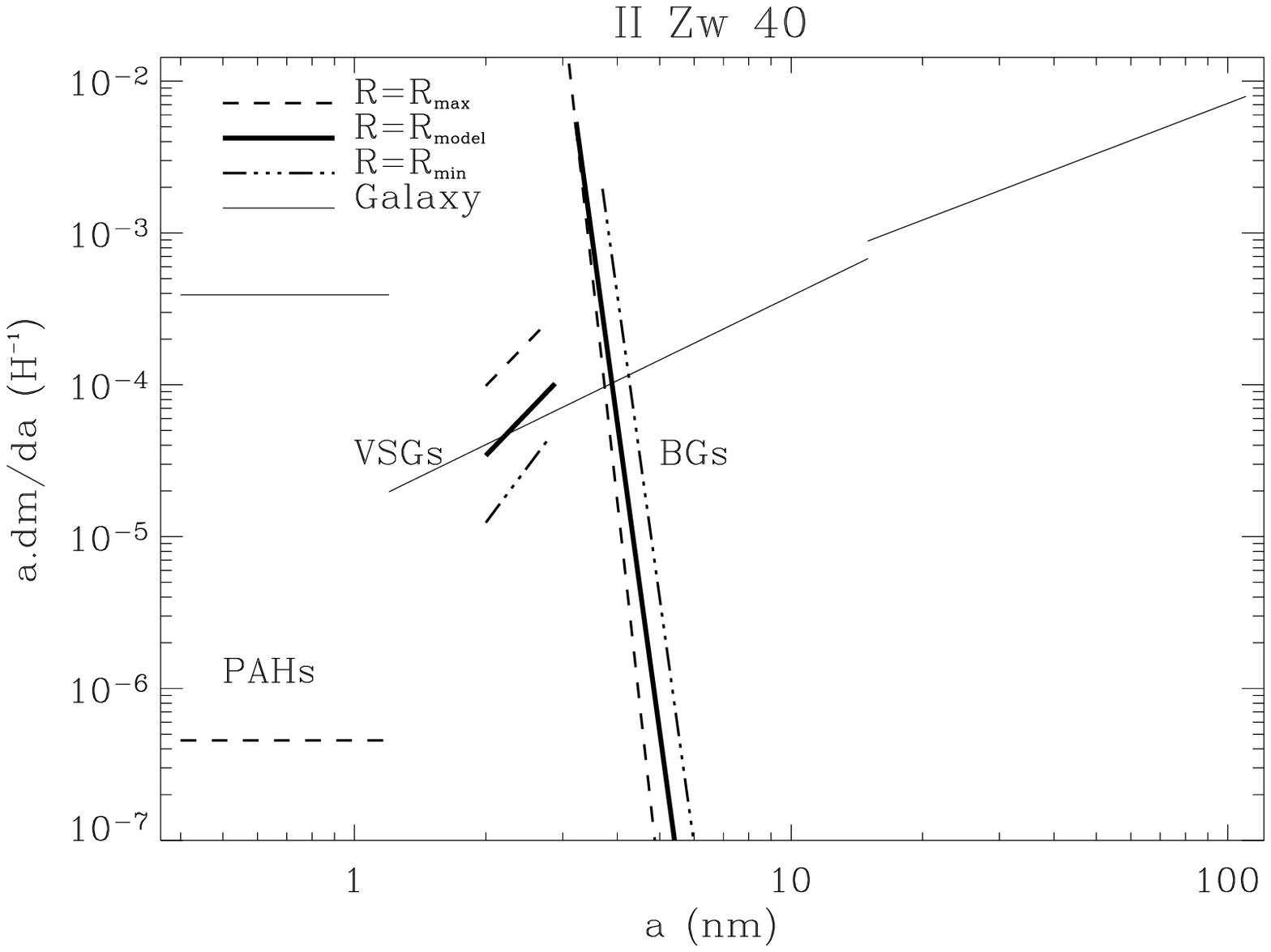} \\
  \includegraphics[width=\linewidth]{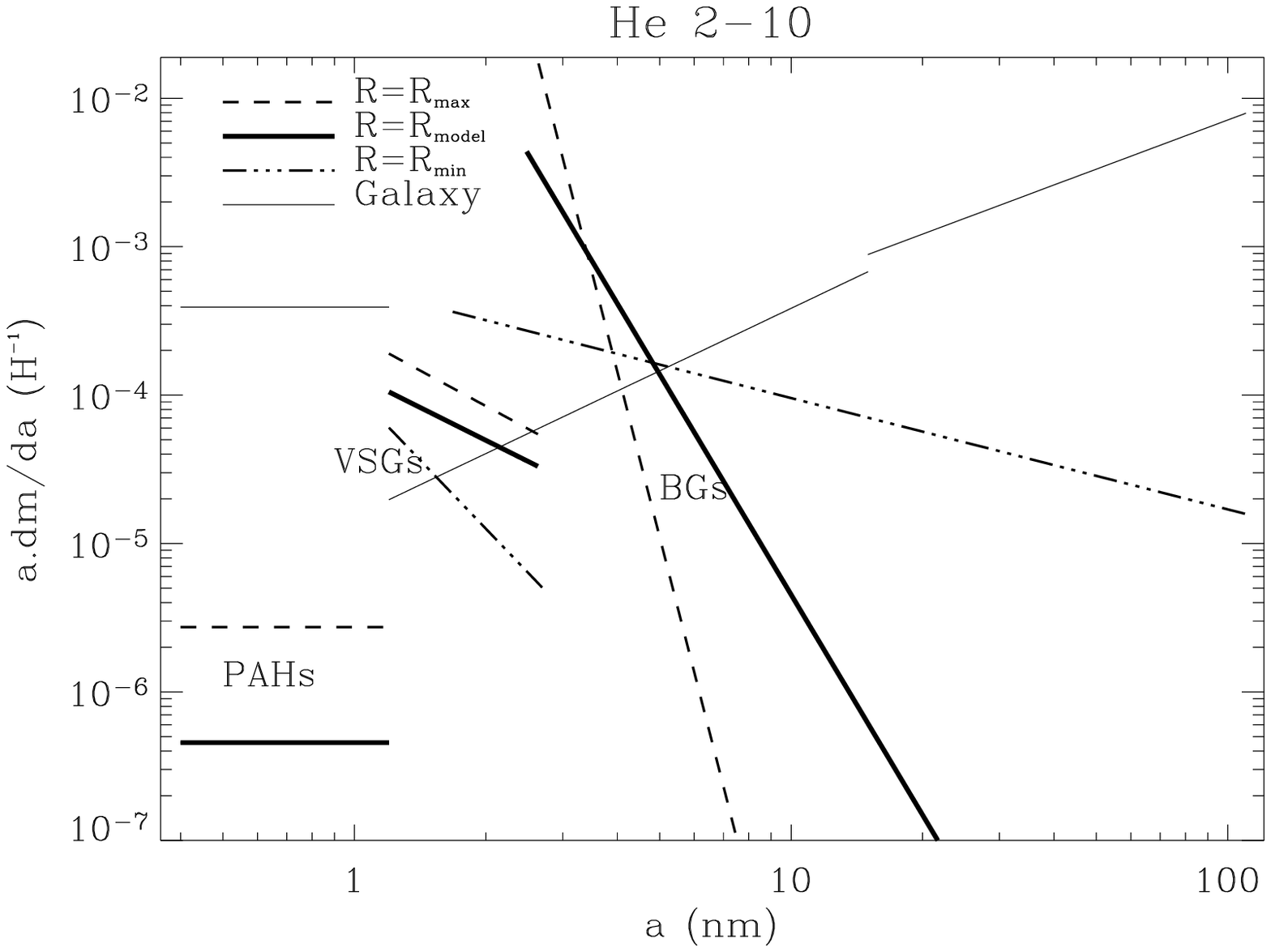} \\
  \includegraphics[width=\linewidth]{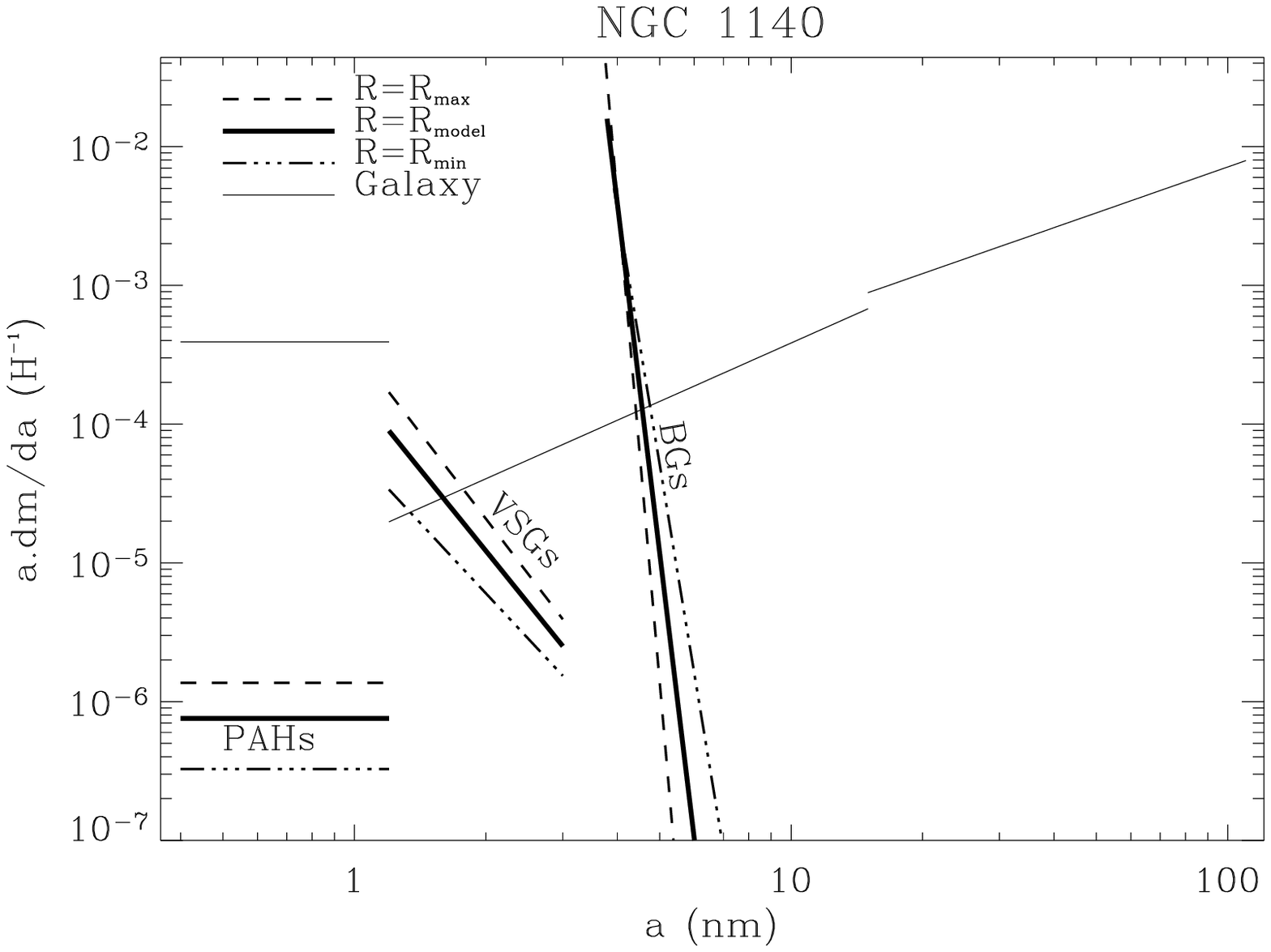}
  \caption{Size distribution deduced for \all.
           The solid thick lines are the best fit solution and the non-solid
           lines are the solutions for the two extreme values of the radius.
           The grey thin line is the size distribution for our Galaxy (\dbp).}
  \label{fig:dist}
  \end{center}
\end{figure}
The size distributions are shown in Fig.~\ref{fig:dist}.
Compared to the Galaxy, there is an obvious dearth of PAHs and the mass 
spectrum is dominated by grains of small sizes (radius $a\simeq 3 - 4\,nm$) 
as shown in Fig.~\ref{fig:dist}. 
The power-law size distributions for all of the galaxies are qualitatively 
similar.
The preference for grains of small sizes may be due to the high supernovae 
rate in these galaxies.
Supernovae produce shocks that fragment and erode the large grains and produce
smaller grains \citep{jones+96}, as we demonstrate in \papii\ where we use the
shock model of \citet{jones+96} to explain the shape of the size distribution 
of \ngc{1569}. 
The effect is similar here for \all.


  \subsection{The modeled SEDs}
  \label{sec:sed}

    \subsubsection{The ISRFs}

\begin{figure}[htbp]
  \centering
  \includegraphics[width=\linewidth]{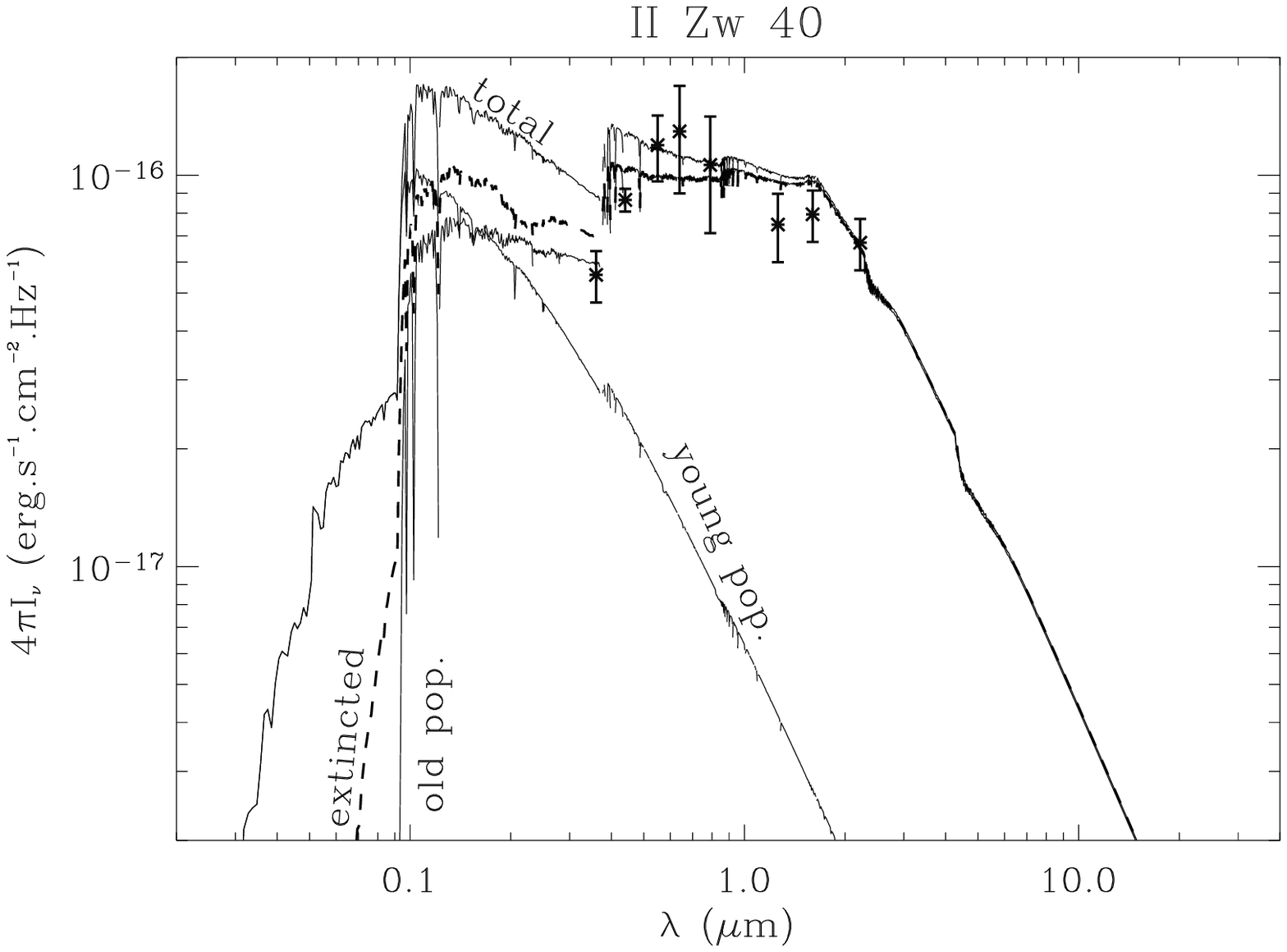} \\
  \includegraphics[width=\linewidth]{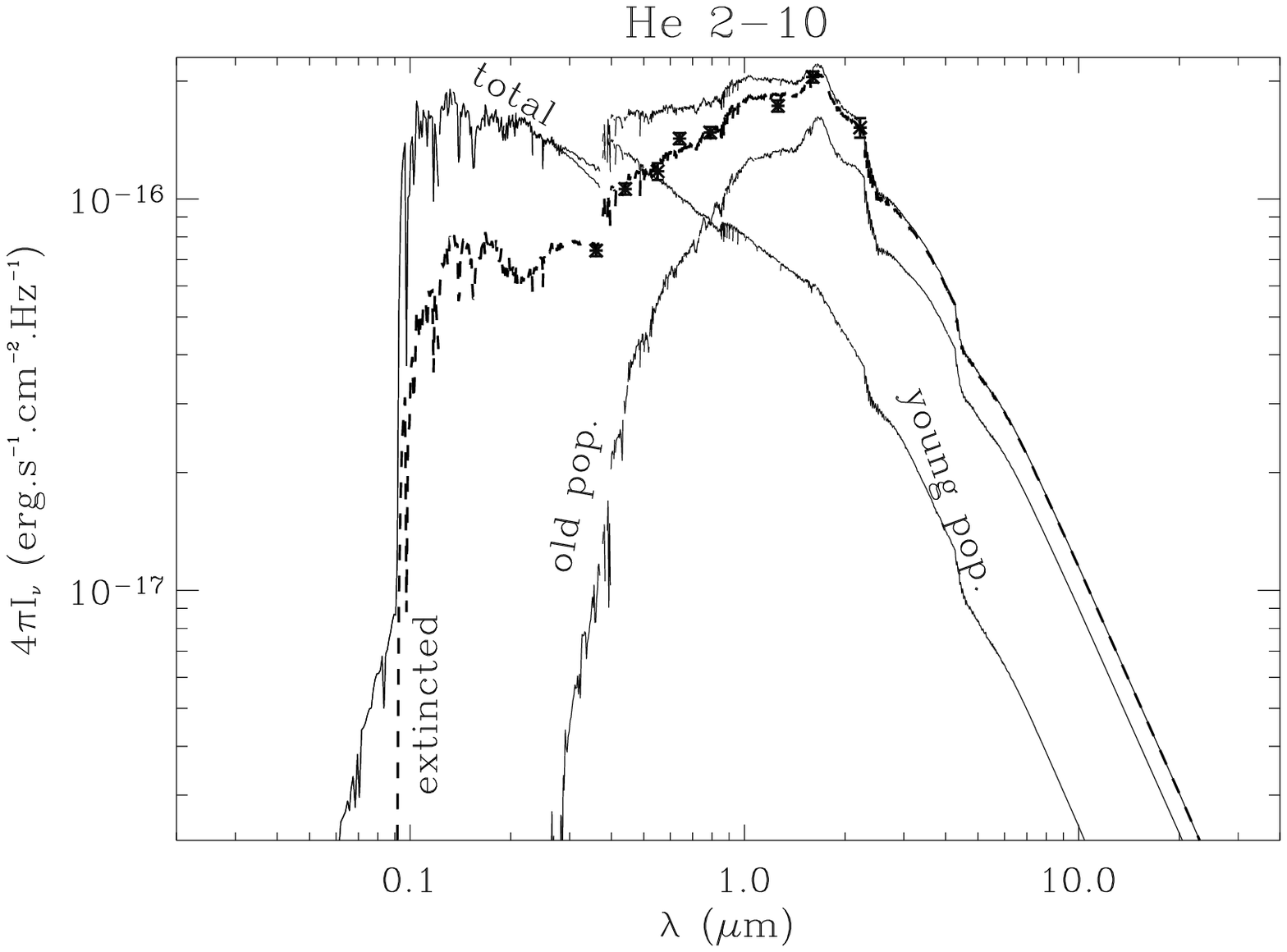} \\
  \includegraphics[width=\linewidth]{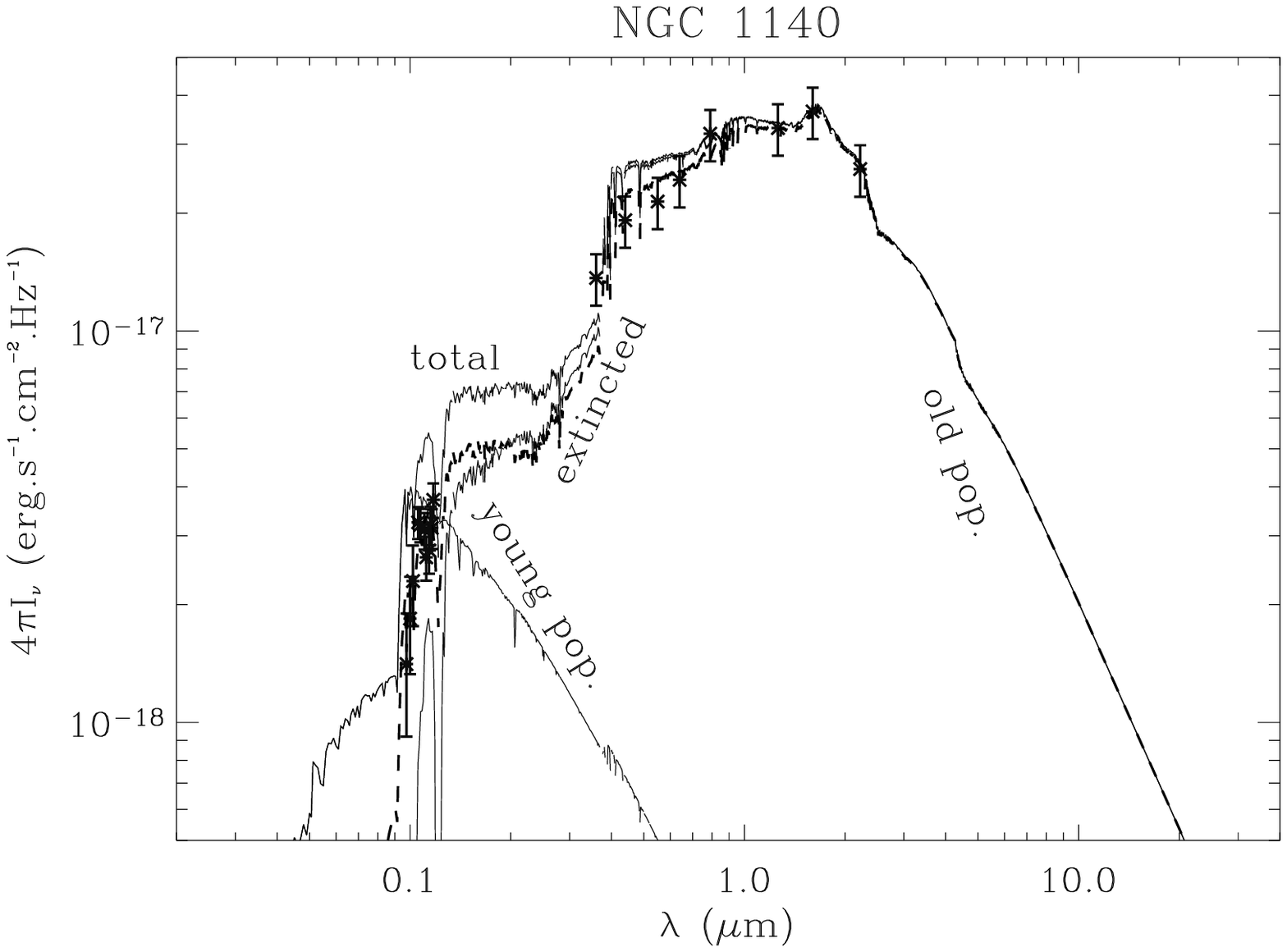}
  \caption{Synthesized ISRFs for \all\ computed with \peg\ and \clo.
           The points with error bars are the observational data from
           Table~\ref{tab:opt}, the solid black line is the global
           non-extincted ISRF, the dashed line is the global extincted ISRF
           and the grey line is the young single-burst component.
           The extinction curve used is the output from the dust model \dbp.
           }
  \label{fig:isrf}
\end{figure}

\begin{figure}[htbp]
  \begin{center}
    \includegraphics[width=\linewidth]{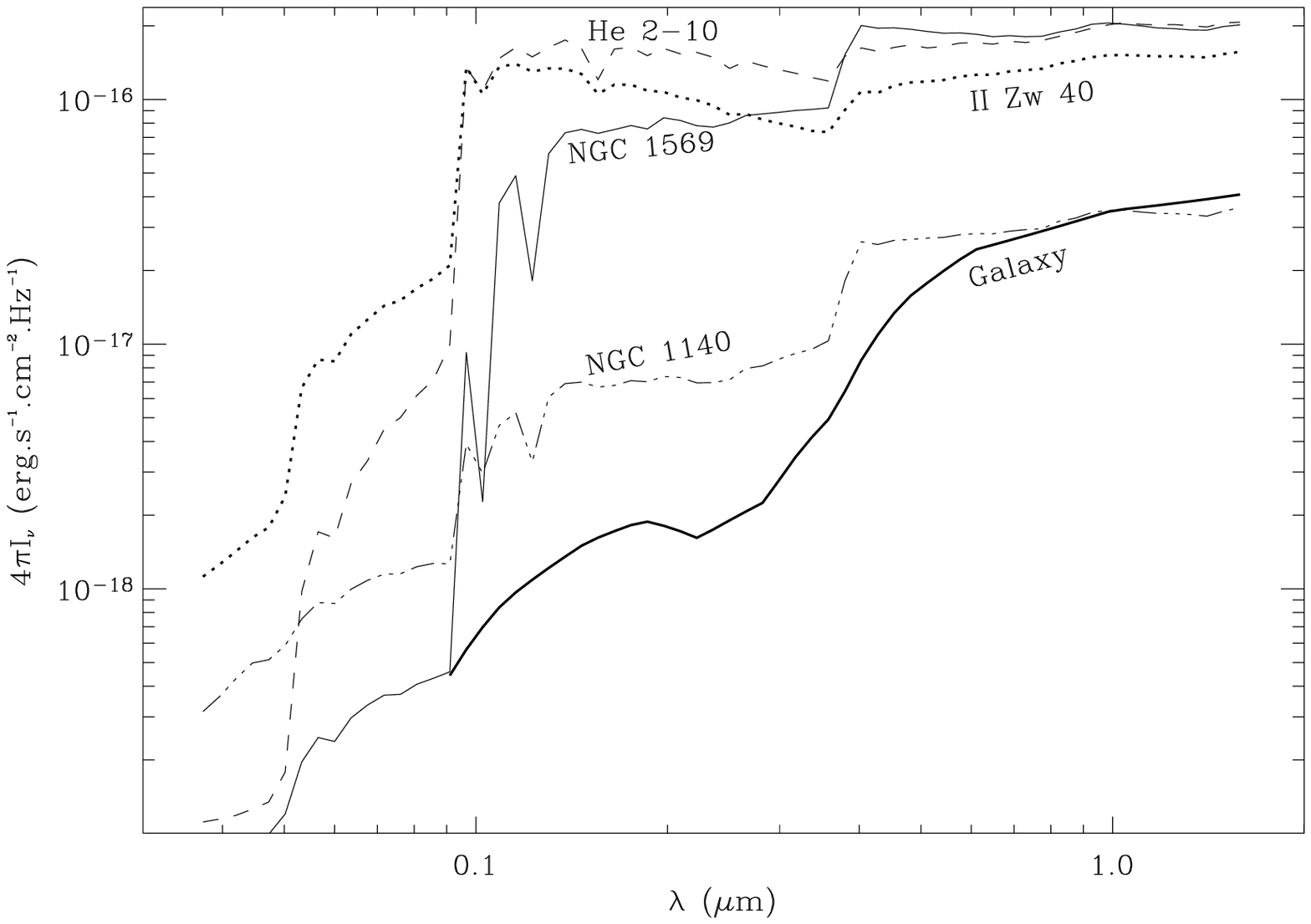}
    \caption{Comparison of the synthesized ISRFs of 
             \ngc{1569} (\papii), \all, with the Galactic one (\dbp).
             These ISRFs are the effective radiation fields seen by
             the dust in our model.}
    \label{fig:compISRF}
  \end{center}
\end{figure}

Fig.~\ref{fig:isrf} shows the synthesized radiation fields for our galaxy 
sample and we compare these with that of the Galaxy in Fig.~\ref{fig:compISRF}.
In the dwarf galaxies with evidence for current starburst activity, the 
effective, global radiation field as seen by the dust, is more intense and 
harder than that of the Galaxy.
In \papi\, we discussed the correlation between the hardness and intensity of 
the ISRF and the lack of PAHs in the starbursting dwarf galaxies, while, PAHs 
are observed to be abundant throughout the Galaxy.
We see an evolutionary sequence among these four galaxies.
\ngc{1140}, where PAHs are evident, is the most quiescent galaxy of the sample,
with a radiation field that is less intense and relatively soft compared to 
the other galaxies (Fig.~\ref{fig:compISRF}; \papi).
In the three other dwarf galaxies, \ngc{1569}, \iizw\ and \hen, PAH bands are 
very weak, if present at all (\papi). 
Likewise, the ISRFs are intense and hard.
Among these three starbursting dwarf galaxies, \ngc{1569} has a softer ISRF 
and the aromatic bands seen in the CVF spectrum are present, even if very weak,
contrary to \iizw, where there is no evidence for the PAH 
band emission in the CVF spectrum.


    \subsubsection{The dust emission}

\begin{figure}[htbp]
  \begin{center}
    \includegraphics[width=\linewidth]{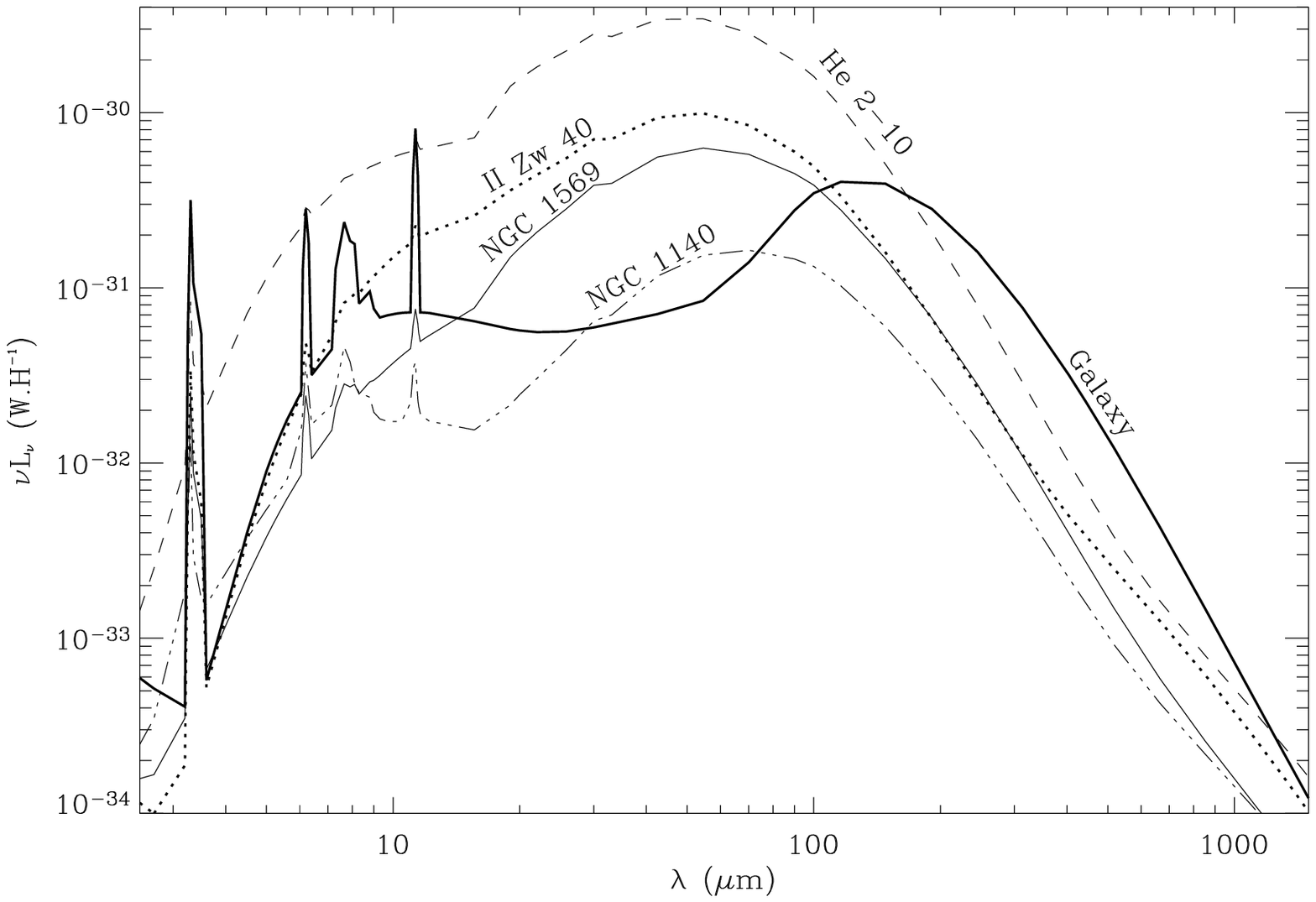}
    \caption{Comparison of the modeled dust SEDs of \ngc{1569} (\papii), 
             \all\ (this paper), and that of the Galaxy (\dbp).}
    \label{fig:compSED}
  \end{center}
\end{figure}

The best $\chi^2$ modeled dust SEDs, after iteration, are plotted in 
Fig.~\ref{fig:sed} and compared to that of the Galaxy in 
Fig.~\ref{fig:compSED}.
The characteristics of the 3 newly-presented dust SEDs have many similarities 
to those already deduced for \ngc{1569}, in \papii.

\begin{figure}[htbp]
  \centering
  \includegraphics[width=\linewidth]{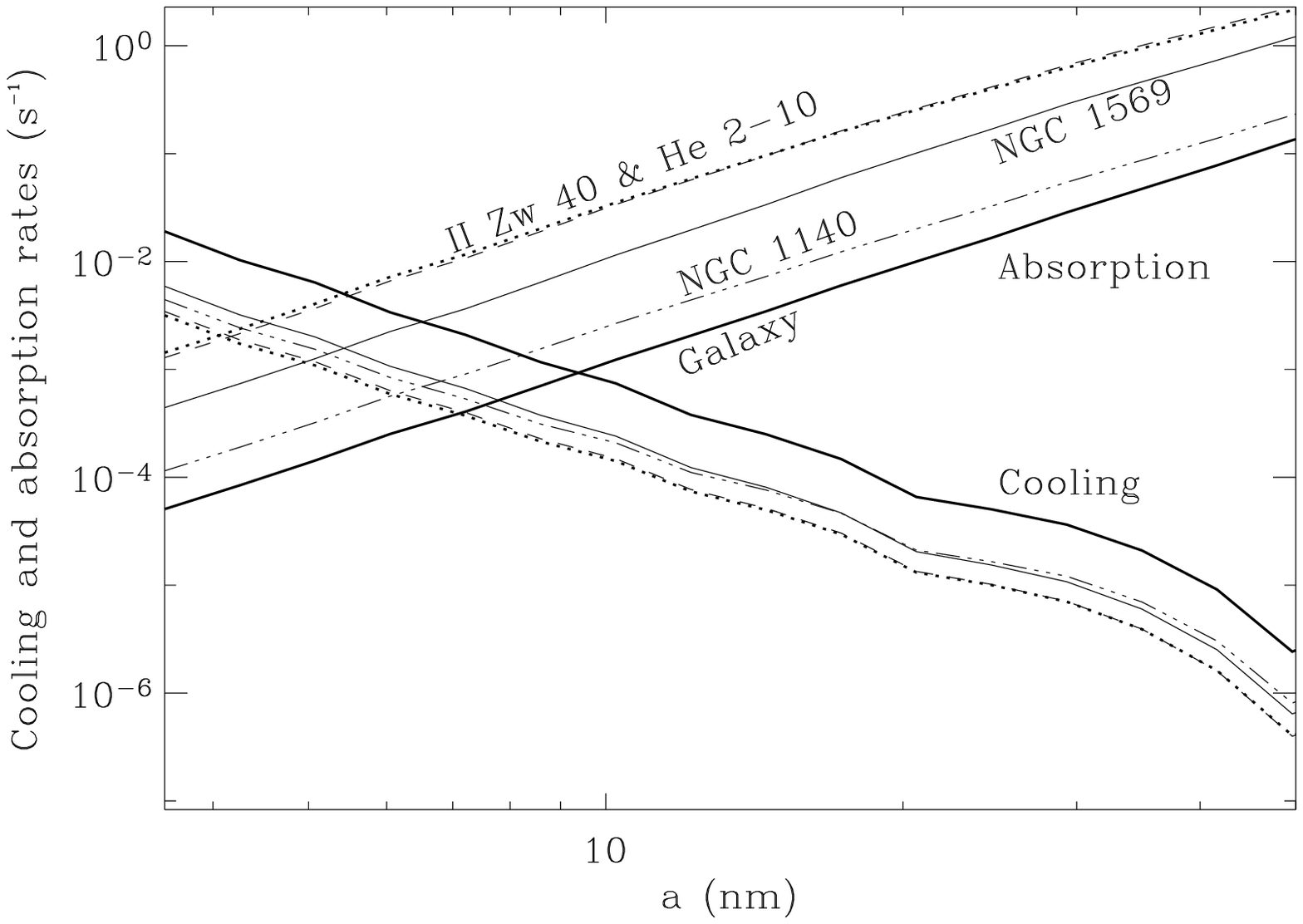}
  \caption{Cooling rates and photon absorption rates as a function of the
           grain radius $a$ for \ngc{1569}, \all, compared to the Galaxy.}
  \label{fig:stochast}
\end{figure}

The SEDs indicate overall hotter dust, peaking at $\sim 60\,\mic$.
Due to the small grain size distributions of the PAH, VSG and BG components, 
most of the grains are stochastically heated and are not in thermal 
equilibrium with the radiation field in these galaxies.
Table~\ref{tab:temp} contains the temperature range of each component for the
maximum and minimum sizes compared to Galactic values.
\begin{table*}[htbp]
  \centering
  \begin{tabularx}{\textwidth}{XX*{5}{|XX}}
  \hline
  \hline
  \multicolumn{2}{l|}{}            & \multicolumn{2}{c|}{\ngc{1569}}
  & \multicolumn{2}{c|}{\iizw}     & \multicolumn{2}{c|}{\hen}  
  & \multicolumn{2}{c|}{\ngc{1140}}    & \multicolumn{2}{c}{Galaxy}
  \\
  \multicolumn{2}{l|}{}            & $T_{\rm min}$ & $T_{\rm max}$
  & $T_{\rm min}$ & $T_{\rm max}$ & $T_{\rm min}$ & $T_{\rm max}$
  & $T_{\rm min}$ & $T_{\rm max}$ & $T_{\rm min}$ & $T_{\rm max}$
  \\
  \hline
  PAH  & $a_-$ 
  & 2.7 K & 9200 K
  & 2.7 K & 9200 K
  & 2.7 K & 9200 K
  & 2.7 K & 9200 K
  & 2.7 K & 4400 K
  \\
       & $a_+$ 
  & 2.7 K & 1800 K
  & 2.7 K & 1800 K
  & 2.7 K & 1800 K
  & 2.7 K & 1800 K
  & 2.7 K & 1100 K
  \\   
  \hline
  VSG  & $a_-$ 
  & 2.7 K & 890 K 
  & 2.7 K & 490 K
  & 2.7 K & 890 K
  & 2.7 K & 890 K
  & 2.7 K & 630 K 
  \\
       & $a_+$ 
  & 2.7 K & 160 K 
  & 2.7 K & 350 K
  & 2.7 K & 380 K
  & 2.7 K & 340 K
  & 2.7 K & 78 K  
  \\   
  \hline
  BG   & $a_-$ 
  & 2.7 K & 230 K 
  & 2.7 K & 140 K
  & 2.7 K & 190 K
  & 2.7 K & 111 K
  & 15 K  & 22 K  
  \\
       & $a_+$ 
  & 28 K  & 28 K  
  & 30 K  & 30 K
  & 31 K  & 31 K
  & 20 K  & 20 K
  & 17 K  & 17 K  
  \\   
  \hline
  \end{tabularx}
  \caption{Temperatures of the PAH, VSG and BG components for \ngc{1569} 
           (\papii), \all\ compared to those of the Galaxy (\dbp).
           We give the minimum ($T_{\rm min}$) and maximum ($T_{\rm max}$)
           temperatures for each component.
           All the grains, except the largest BGs
           (for which $T_{\rm min} = T_{\rm max} = T_{\rm equilibrium}$), 
           are stochastically heated.
           The values of $T_{\rm max}$ given for the PAHs are larger than the 
           vaporisation temperature, thus they are unphysical.
           The minimum temperature never reaches values below 
           $T_\sms{CMB}$, since all grain sizes are in equilibrium with
           the very low energy input coming from the CMB.}
  \label{tab:temp}
\end{table*}
In \papii, we estimated the transition grain radius, $a_t$, between stochastic
heating and thermal equilibrium to have an idea over what range of dust sizes 
it is important to consider the process of stochastic heating. 
This is a function of the radiation field, the dust size and the heat capacity
of the dust.
We refer to Section~4.1.3 of \papii\ for a detailed explanation of the 
calculation.
In Fig.~\ref{fig:stochast}, we compare the variation of the cooling rates and 
the photon absorption rates as a function of the radius, for a single BG.
The point where the two curves intersect gives an idea of the grain radius, 
$a_t$, below which the stochastic heating becomes dominant.
These transition radii are $a_t\simeq 5\,$nm in \ngc{1569},
$a_t\simeq 4\,$nm in \iizw\ and \hen, $a_t\simeq 6\,$nm in \ngc{1140} and 
$a_t\simeq 9\,$nm in the Galaxy.
These estimates demonstrate that the grains become stochastically heated at 
smaller radii, due to the higher intensity of the radiation field
(Fig.~\ref{fig:compISRF}).
By comparing the estimates of $a_t$ with the typical radius that dominates 
the mass spectrum of the grains, which is $a\simeq 3-4\,$nm 
(Fig.~\ref{fig:dist}), we deduce that most of the grains are stochastically 
heated in these galaxies.
Even at wavelengths as long as $60\,\mic$, where the dust emission peaks, the 
grains are primarily stochastically heated.

Another striking feature of the dust emission spectra is the weakness or 
absence 
of the \mir\ PAH emission bands particularly in \ngc{1569} and \iizw\ (the 
\mir\ spectra are presented in \papi).
\hen\ has not been observed with the ISOCAM CVF. 
However, \citet[\privcom\ from Marc Sauvage]{martin-hernandez+05} have 
obtained a ground-based \mir\ spectrum in the
range $8-13\,\mic$ which also reveals a lack of PAH features.
Therefore, for these three galaxies, we only put an upper limit on the PAH 
mass, given by the continuum emission of this component. 
We do not actually fit the bands in the model.
In the case of \ngc{1140}, we see relatively significant PAH emission, 
allowing us to properly constrain the PAH mass.
For this purpose, we use a modified version of the \dbp\ model (Laurent 
Verstraete, \privcom) where the PAH emission features are modeled more 
precisely (Fig.~\ref{fig:sed}).
The optical constants are deduced from a spectrum of a typical \hii\ region.
For \ngc{1140}, we use the process of decomposition of the ISOCAM \mir\ 
spectrum described in \papi: the PAH features are modeled as Lorentzian bands;
the ionic lines are modeled as Gaussian emission lines and the continuum is 
fitted as a modified black body providing a constraint on the VSG component in
the dust model.
Then, we add the PAH part of the model to fit the entire spectrum, including 
the aromatic bands.

For each SED of our sample, we are left with a submillimetre/millimetre 
excess that we cannot explain with the standard \dbp\ model.
As originally found for \ngc{1569}, we are compelled to invoke the presence 
of an ubiquitous very cold grain component (VCGs) in \all. 
Various hypotheses for this excess are fully explored in \papii. 
They include: very cold dust, change of grain optical properties at long 
wavelengths and grain-grain coagulation.
We modeled the VCG component using a modified black body 
(Sect.~\ref{sec:param}).
The temperatures of these VCG components are $T = 5 - 9\,$K 
(Table~\ref{tab:paramerr}).
We have shown in \papii\ (Sect.~4.5) that the VCG component could correspond
to very cold dust embedded in very dense clumps.
Other means to increase the submillimetre emissivity which we see in
these galaxies could be different optical properties due to
temperature effects, as well as grain-grain coagulation. 
The temperature dependent optical properties are considered in \papii\ 
and in the Appendix~\ref{app:agladze} of the present paper, but 
so far failed to produce better fits. 
Thus, in \papii\ , as well as the present paper, we chose to discuss the very
cold dust hypothesis, since it has a number of quantifiable
consequences (Sect.~\ref{sec:clumps}).


  \subsection{The extinction curves}
  \label{sec:ext}

\begin{figure}[htbp]
  \begin{center}
    \includegraphics[width=\linewidth]{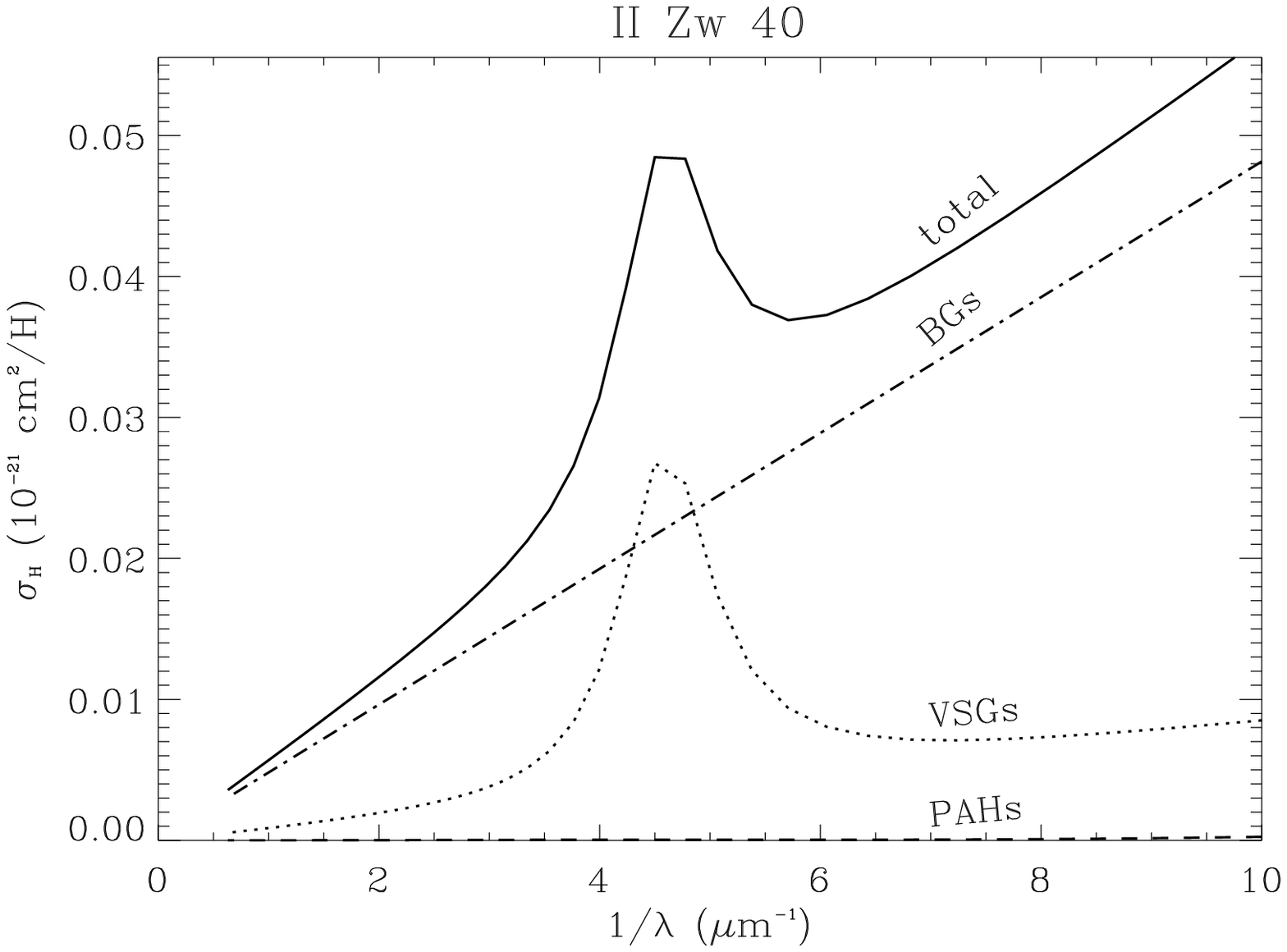}  \\
    \includegraphics[width=\linewidth]{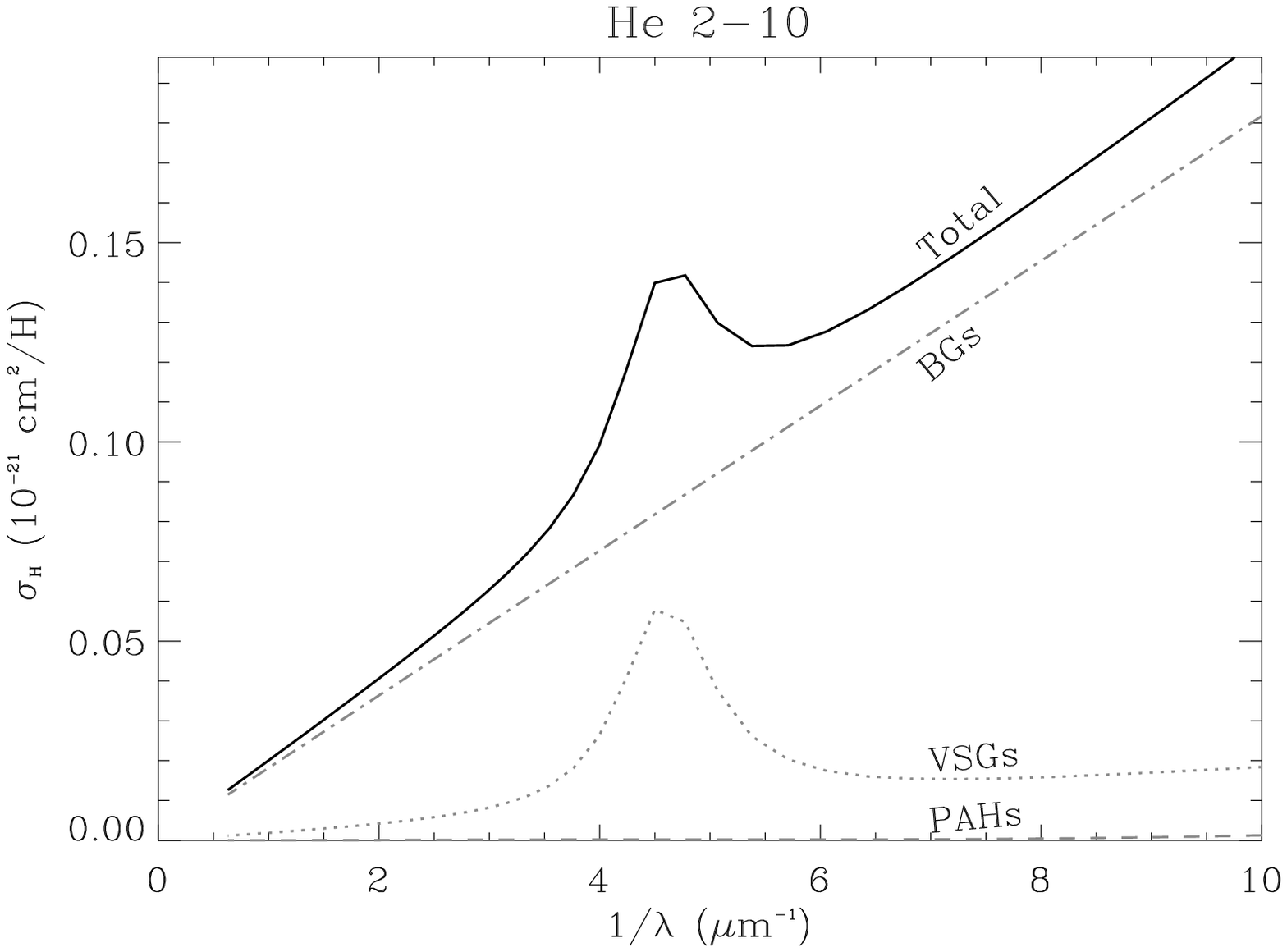}  \\
    \includegraphics[width=\linewidth]{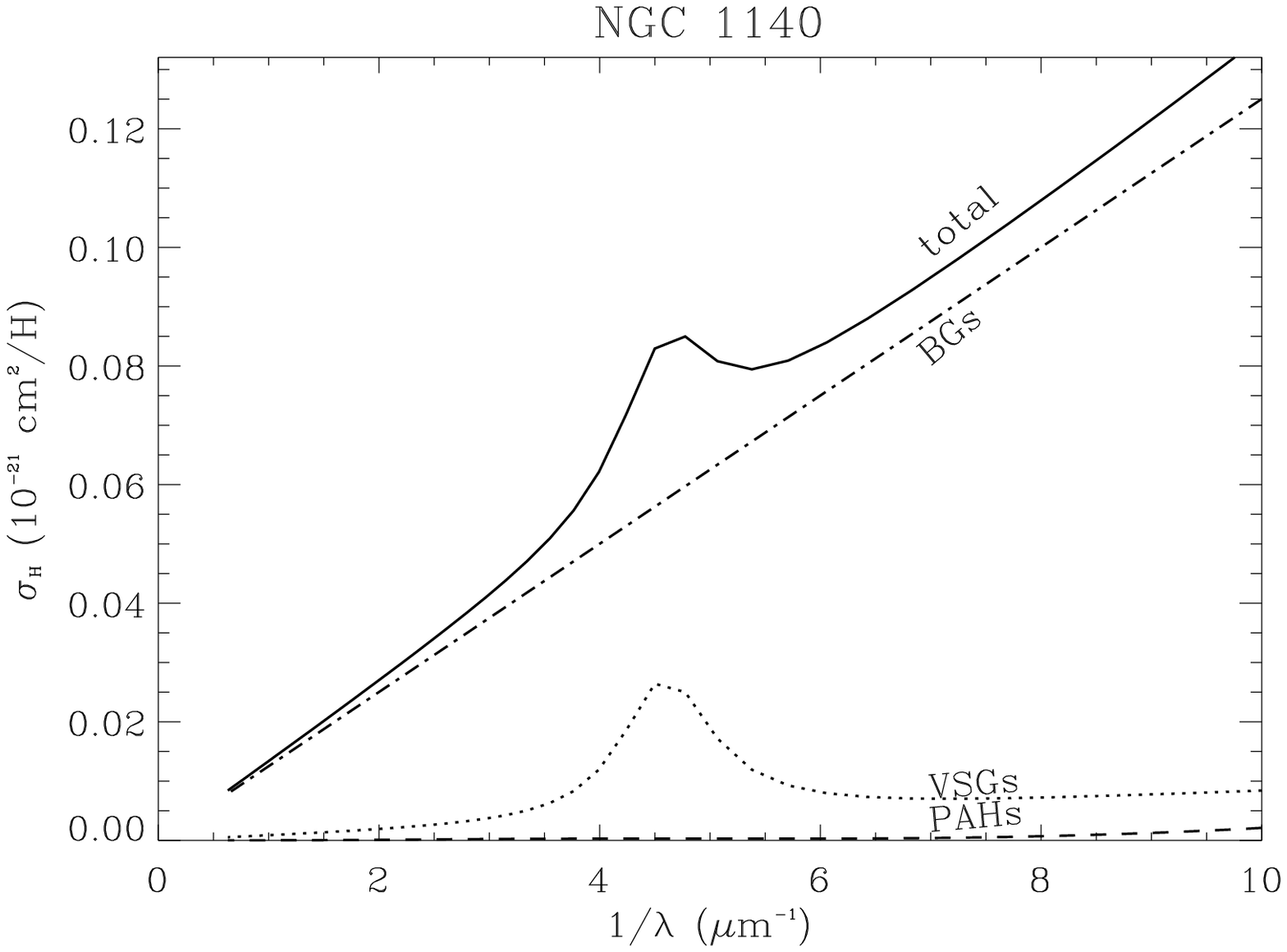}
    \caption{Synthesized extinction curves for \all. 
             The contributions to the extinction are shown individually for 
             the PAH, VSG and BG components.
             The solid lines are the total extinction curves modeled with \dbp.
             The opacity is expressed as the cross section per H atom.}
    \label{fig:ext}
  \end{center}
\end{figure}

\begin{figure}[htbp]
  \begin{center}
    \includegraphics[width=\linewidth]{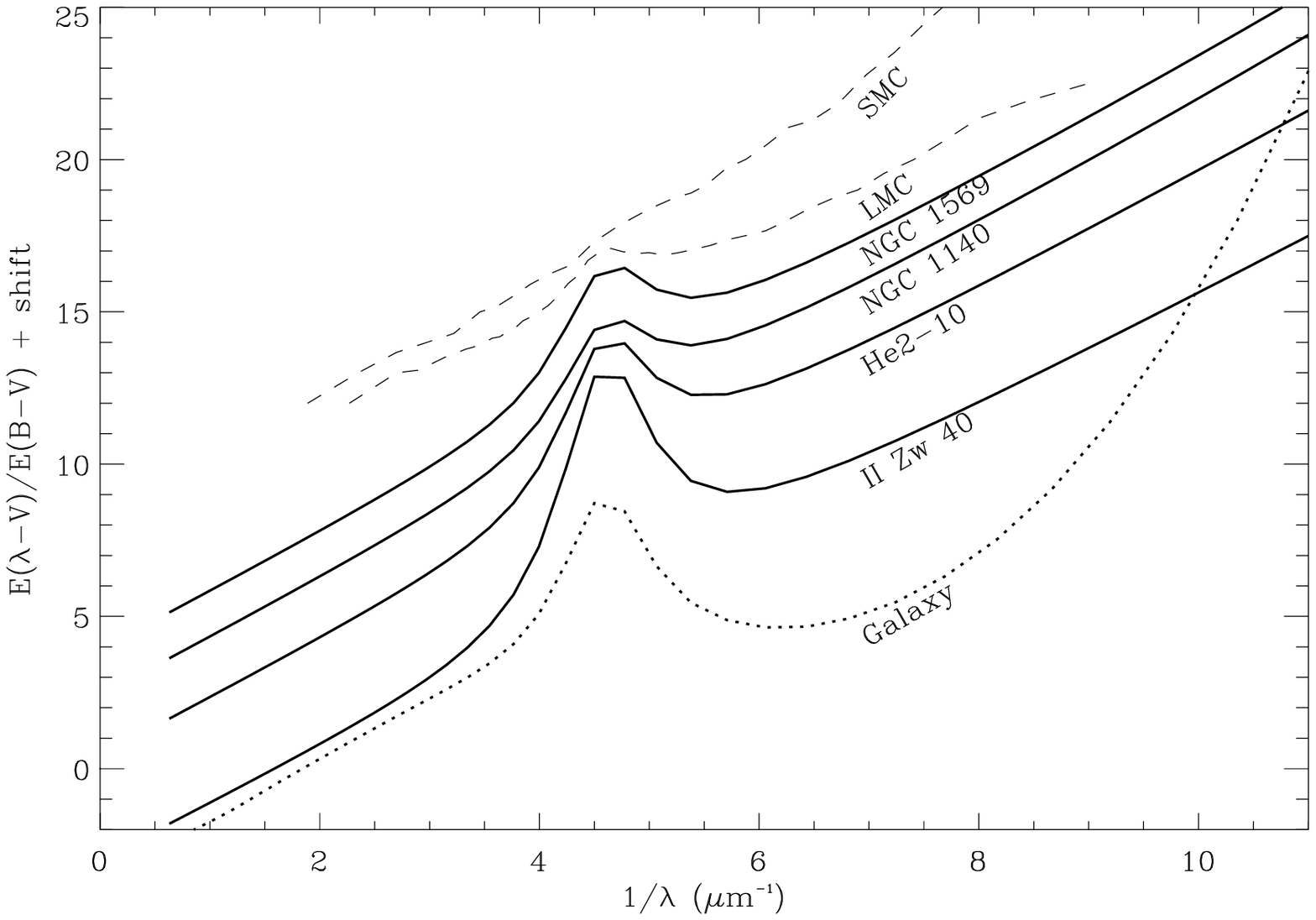}
    \caption{Comparison of the synthesized extinction curves of 
             \ngc{1569} (\papii), \all\ with the Galactic extinction curve 
             (\dbp), the observed LMC curve (average of \citet{koorneef+81}
             and \citet{nandy+81}) and the observed SMC curve 
             \citep{prevot+84}.
             These curves have been shifted for clarity.}
    \label{fig:compExt}
  \end{center}
\end{figure}

From the dust properties described by the parameters in 
Tables~\ref{tab:paramod} and \ref{tab:paramerr}, we synthesize an extinction 
curve assuming a simple screen attenuation of the radiation. 
Construction of the extinction curve is based on the \dbp\ model assumptions 
of the PAHs being the carriers of the FUV non-linear rise; 
the carbonaceaous VSGs giving rise to the 2175$\,$\AA\ extinction bump 
\citep[{\eg}][]{savage+79} and the BGs giving rise to the NIR and visible rise 
of the extinction curve.
The individual contributions to the extinction curves and the total 
synthesized extinction curves are displayed in Fig.~\ref{fig:ext} and are
compared to the \mw\ (\dbp), the Large Magellanic Cloud 
\citep{koorneef+81,nandy+81} and one line of sight toward the Small Magellanic 
Cloud \citep{prevot+84} in Fig.~\ref{fig:compExt}. 

These synthesized global extinction curves are characterised by a 
quasi-linearity with $1/\lambda$ which is also the case for the extinction 
curves seen toward the Magellanic Clouds (Fig.~\ref{fig:compExt}).
These slopes indicate a higher absorption in the UV as a consequence of the 
presence of the small sizes of the BGs, as can be seen on the decompositions 
in Fig.~\ref{fig:dist}, and also due to the lack of PAHs which are responsible
for the \fuv\ rise.
We also note that the 2175$\,$\AA\ extinction bump is lower in \ngc{1569}, 
\hen\ and \ngc{1140} than in the \mw, as is the case for the extinction curves
observed toward a line of sight in the low-metallicity Magellanic Clouds.
This is a consequence of the higher silicate-over-graphite ratio 
($M_\sms{BG}/M_\sms{VSG}$), in these galaxies than in the \mw\ (this ratio is 
equal to 14 for the \mw, and to 24, 18 and 28 for \ngc{1569}, \hen\ and 
\ngc{1140}, respectively, from Table~\ref{tab:paramod}).
Indeed, in the \dbp\ model, the carriers of the bump are the VSGs. 
However, the ratio $M_\sms{BG}/M_\sms{VSG}$ is smaller in \iizw\ 
($M_\sms{BG}/M_\sms{VSG}\simeq11$), which explains why the model determines a 
larger bump in \iizw. 
The silicate-graphite ratio is not constrained very accurately
as can be seen from the error bars on our parameters in
Table~\ref{tab:paramerr}.
Moreover, the carriers of the bump could be other species of grains like the
PAHs, as proposed by \citet{joblin+92}.
In recent dust models \citep{draine+01,li+01,zubko+04}, the carriers of the 
bump are both PAHs and solid carbonaceous grains.
Thus, the lack of PAHs could naturally explain the weakness of the bump in 
these galaxies.

We think, as already suggested \eg\ by \citet{mas-hesse+99} and 
\citet{weingartner+01}, that the shape of the extinction curve, thus the very 
different dust properties, essentially reflects the starburst activity in 
these galaxies, rather than a consequence of their low metallicity.
Indeed, the fragmentation and the erosion of the grains by the numerous
shock waves following the starburst, as suggested here, could be the 
main process determining the extinction properties, rather than a 
deficit in formation of large grains by evolved stars.
This argument is also supported by the fact that observations toward some 
lines of sight in the SMC show Galactic type extinction 
\citep{gordon+98,gordon+03} and observations toward some lines of sight 
inside the Galaxy show SMC like extinction curves \citep{valencic+03}.


  \subsection{The dust masses}
  \label{sec:mass}

Table~\ref{tab:mass} contains all of the most interesting physical quantities
deduced from our modelling. 
The dust masses are directly deduced from Tables~\ref{tab:paramod} and
\ref{tab:paramerr}.
Table~\ref{tab:mass} highlights the fact that most of the mass is contained
in the VCG component, if we believe that the submillimetre excess is produced
by very cold dust.
There is a large uncertainty on the mass of this component since its 
temperature is uncertain.
However, our results show that between 40 and 80$\,\%$ of the dust mass is
in the form of VCGs. 
The rest of the dust mass is dominated by the BGs.
The VSGs, while the most numerous particles, contain between 1 and 3$\,\%$ of 
the total dust mass.
\begin{table*}[htbp]
  \centering
  \begin{tabularx}{\linewidth}{*{5}{X}}
    \hline
    \hline
                                 & \ngc{1569}                   
    & \iizw\                         & \hen\ 
    & \ngc{1140}                        \\
    \hline
    $L_{\rm dust}$ ($\lsol$)     & $(5.8\pm 0.1)\times 10^8$ 
    & $(2.0\pm 0.2)\times 10^9$      & $(5.4\pm 0.2)\times 10^9$
    & $(3.3\pm 0.1)\times 10^9$      \\ 
    $L_\star^{\rm ext}$ ($\lsol$)& $(1.3\pm 0.1)\times 10^9$ 
    & $(2.5\pm 0.1)\times 10^9$      & $(5.3\pm 0.2)\times 10^9$
    & $(6.0\pm 0.2)\times 10^9$      \\
    $\tau_V^{\rm eff}$           & 0.24             
    & 0.15                           & 0.26
    & 0.11                           \\
    \hline
    $M_{\rm PAH}$ ($\msol$)      & $\lesssim 190$                 
    & $\lesssim 220$                 & $\lesssim 930$
    & $6.2^{+4.2}_{-2.5}\times 10^3$ \\
    $M_{\rm VSG}$ ($\msol$)      & $3.4^{+1.0}_{-0.9}\times 10^3$
    & $1.0^{+0.7}_{-0.4}\times 10^4$ & $1.5^{+0.8}_{-0.7}\times 10^4$
    & $1.7^{+0.9}_{-0.4}\times 10^5$ \\
    $M_{\rm BG}$ ($\msol$)       & $8.4^{+0.5}_{-1.7}\times 10^4$
    & $1.1^{+0.6}_{-0.3}\times 10^5$ & $2.8^{+1.9}_{-1.2}\times 10^5$
    & $4.6^{+2.3}_{-2.0}\times 10^6$ \\
    $M_{\rm VCG}$ ($\msol$)      & $(0.7 - 2.5)\times 10^5$
    & $(3.1 - 8.8)\times 10^5$       & $(0.2 - 1.3)\times 10^6$
    & $(0.2 - 1.3)\times 10^7$       \\
    Total $M_d$ ($\msol$)        & $(1.6 - 3.4)\times 10^5$
    & $(0.4 - 1.1)\times 10^6$       & $(0.4 - 1.8)\times 10^6$
    & $(0.5 - 2.0)\times 10^7$       \\
    \hline
    $\mathcal{G}$                & $740 - 1600$
    & $530 - 1460$                   & $330 - 1500$
    & $500 - 2000$                   \\
    $\mathcal{D}$                & $1/6 - 1/3$
    & $1/5 - 1/2$                    & $1/30 - 1/6$
    & $1/13 - 1/3$                   \\
    \hline
  \end{tabularx}
  \caption{Derived quantities from our dust modelling, for \ngc{1569} 
           (\papii), \all.
           $L_{\rm dust}$ is the luminosity reemitted by the dust from \nir\ 
           to mm.
           $L_\star^{\rm ext}$ is the escaping stellar luminosity.
           $\tau_V^{\rm eff}$ is the effective optical depth deduced from the 
           energy balance (see \papii).
           $M_{\rm PAH}$, $M_{\rm VSG}$, $M_{\rm BG}$ and $M_{\rm VCG}$ are the
           masses of each dust component and $M_d$ is the total dust mass.
           $\mathcal{G}$ is the gas-to-dust mass ratio and $\mathcal{D}$ is
           the dust-to-metal mass ratio. 
           The Galactic values are typically 
           $\mathcal{G} \simeq 150$ and $\mathcal{D} \simeq 1/3$.
           }
  \label{tab:mass}
\end{table*}
To compute the gas-to-dust mass ratio we assume that 
$M_{gas} = M(\mbox{\hi}) + M({\rm H_2}) + M({\rm He})$ and that 
$M({\rm He}) = 0.25\times M_{gas}$.
The $M(\mbox{\hi})$ are taken from table~\ref{tab:physpar} and 
$M({\rm H_2})$ are given in the literature:
\citet{meier+01} report $M({\rm H_2}) < 0.4\times 10^6\;\msol$ for \iizw\ 
and $M({\rm H_2}) = 1.4\times 10^8\;\msol$ for \hen.
For \ngc{1140}, we assume that $M({\rm H_2})$ is negligible compared to 
$M(\mbox{\hi})$, since we did not find any global estimate in the literature
and this mass is likely to be small.
We use the following definition of the dust-to-metal mass ratio:
$\mathcal{D} = 1/\mathcal{G}Z$, which is the fraction of metals locked up in 
dust.
The Galactic values are typically $\mathcal{G} \simeq 150$ and 
$\mathcal{D} \simeq 1/3$.
The range of values we give for $\mathcal{G}$ and $\mathcal{D}$ are rather 
large (Table~\ref{tab:mass}).
The constraints that we put on this parameter allows lower values of the 
dust-to-metal mass ratio than established by other studies 
\citep{lisenfeld+98}.
The dust-to-metal ratio for the dwarf galaxies may be lower than that 
of the \mw, except for the upper limit of \iizw.
It would suggest that low-metallicity systems could be less efficient in 
forming dust than normal metallicity galaxies.

\hen\ is a special case, since the VCG mass we computed is affected by the fact
that the upper limit of the free-free emission we subtracted from the 
millimetre fluxes in order to be conservative, is significant, and also, 
since its metallicity is relatively uncertain.
This is the reason why the value of $\mathcal{D}$ in \hen\ deviates from the 
others.


  \subsection{The nature of the very cold grain component (VCGs)}
  \label{sec:clumps}

In \papii, we discussed the likeliness of the VCG hypothesis by estimating
order of magnitudes of the temperature and the filling factor of dust
hidden in very dense clumps. 
Here, we do the same calculation for \all.

Following the equation~(11), in \papii, we compute a rough estimate of the
temperatures of the VCGs, $T_{\rm VCG}$.
This is the temperature that would be reached by grains totally shielded from 
the stellar radiation, heated only by the emission from the surrounding dust.
It is estimated assuming that the inferred mass of VCGs will be heated only by
the radiation arising from the dust peaking in the FIR in our synthesised SEDs.
The ranges of values that we calculate are: 
$T_{\rm VCG}\simeq 7 - 9\,$K for \iizw, 
$T_{\rm VCG}\simeq 8 - 10\,$K for \hen\ and
$T_{\rm VCG}\simeq 6 - 9\,$K for \ngc{1140}.
All these values are consistent with the temperatures given in 
Table~\ref{tab:paramod} except in the case of \hen\ where the high level of
the upper limit of the free-free emission gives more uncertainty on the VCG
parameters.
We believe that these figures support the hypothesis of very cold dust in 
dense clumps to explain the submm excess of the SEDs.

We are also able to give an idea of the size of the clumps and of their 
filling factor.
As we did in \papii, for \ngc{1569}, we assume that the BGs are shielding the 
VCGs.
If we assume typical contrast densities of $\delta = 10^4 - 10^5$, we find
clump sizes of 
$0.9 - 18\,$pc for \iizw,
$1.2 - 38\,$pc for \hen\ and
$2.4 - 43\,$pc for \ngc{1140}.
The corresponding filling factors are:
$2\times 10^{-4} - 2\times 10^{-3}$ for \iizw,
$3\times 10^{-4} - 3\times 10^{-3}$ for \hen\ and
$4\times 10^{-5} - 8\times 10^{-4}$ for \ngc{1140}. 
Thus, our conclusion is that the ISM of these galaxies could be very clumpy 
with low filling factors.
In contrast, the filling factor of the Galactic molecular phase is about 
$1\,\%$ \citep{tielens95}.
Recently, \citet{andre+04}, found very small ($\simeq100\,$pc) dense CO clumps 
in the low-metallicity ISM of the Magellanic clouds.
These clumps could contain the very cold dust probed by our submm-mm SEDs.
\citet{rubio+04} found that the CO mass deduced from the 
submillimeter continuum emission of molecular cloud in the SMC, is $\sim 10$ 
times higher than the virial mass reported from CO observations.
Their interpretation is that the CO emission comes from dense clumps.


\section{Summary and conclusion}
\label{sec:concl}

We have presented new SCUBA images at $450\,\mic$ and $850\,\mic$ and
MAMBO ON-OFF observations of \all.
With additional data from the literature, we have constructed the observed
SEDs for these galaxies.
Using the same modelling approach as in \papii, we have combined
stellar evolution, photoionisation and dust modelling to compute 
self-consistent dust SEDs.
The results found in \papii, for \ngc{1569}, are supported here, for \all.
\begin{enumerate}
  \item We find very low abundances of PAHs, and smaller overall sizes 
  ($\sim 3 - 4\,$nm) of grains emitting in the \mir\ and \fir, in contrast to 
  grains observed in the Galaxy.
  The small sizes of the grains, on average, are supported by the erosion
  by shock waves produced by the numerous supernovae that occured in these
  starburst galaxies.
  \item Due to the small sizes of the grains, the stochastic heating of dust
  is predominant in these galaxies, even the grains emitting at \fir\ 
  wavelengths.
  The bulk of the emission comes from grains which are not in thermal
  equilibrium with the radiation field.
  \item In each of our four SEDs presented here and in \papii, we find, with
  no exception, a submillimetre emission excess.
  We propose that this excess is the emission of very cold dust 
  ($5\,{\rm K}\lesssim T \lesssim 9\,$K) hidden in dense clumps.
  In each of these SEDs, except one, we find that $\beta = 1$ is the most 
  likely value for the emissivity index of this component, even if this 
  parameter is difficult to accurately constrain with the scarcity of data
  in the submillimetre-millimetre part of the spectra.
  Very cold grains represent between 40 and 80$\,\%$ of the total dust mass.
  We show that this very cold dust hypothesis is consistent, although we can
  not exclude other explanations for this millimetre excess, like non-standard
  optical properties.
  \item The submillimetre-millimetre emission is not concentrated toward the
  outer regions of the galaxies but correlated with the \mir\ emission 
  distributed around the star forming regions.
  Our results are consistent with a clumpy medium composed of ubiquitous
  clumps of sizes between a few pc to a few tens of pc with typical filling 
  factor of $10^{-4} - 10^{-3}$.
  \item The extinction curves that we synthesize for the 4 dwarf galaxies are 
  different from that of the Galaxy and similar in shape to that of the LMC.
  Due to the small sizes of the grains and the low abundance of PAHs (or lack 
  of PAHs), the slope of the extinction curves are constant from \nir\ to 
  \fuv\ wavelengths with the greatest extinction in the UV range.
  The extinction bump at 2175$\,$\AA, which is assumed to be carried by the 
  very small grains in the dust model we use, is found to be smaller than the 
  Galactic one in our sample except in one case where we have a high \mir\ 
  attenuation.
  \item We constrain the gas-to-dust mass ratios which lie between 300 and 2000
  and the dust-to-metal mass ratios which lie between $1/30$ and $1/2$ in 
  our sample.
\end{enumerate}

This paper closes a serie of three publications related to the MIR to mm 
properties of low-metallicity environments. 
This study investigates the dust properties of 3 new low-metallicity galaxies,
\all\ and along with \ngc{1569} (\papii), presents the first detailed SED 
studies focused on dwarf galaxies.
The study of dust properties using IR emission provides results that are 
overlooked  by optical studies based on the extinction phenomenon.
Dust properties are profoundly different in these galaxies compared to our 
Galaxy and other normal galaxies.
The dust size distribution is different, consequently the SEDs and the
extinction laws are not the same.
Moreover the dust mass and temperature estimates are affected by both the 
presence of very cold dust and by the non-thermal equilibrium heating of the 
bulk of the emitting dust.
Assuming Galactic dust properties when looking at these galaxies is not 
correct and could lead to large errors, particularly in dust masses.

We also warn the reader that it is very important to obtain data in
the MIR regime as well as in the submillimetre/millimetre regime to
properly constrain the dust properties.  
The submillimetre/millimetre regime is still largely unknown and poorly 
sampled.  
Observations of galaxies at these wavelengths could contribute significantly 
to the comprehension of the variety of galactic dust environments.  
Without such precise dust SED modeling incorporating the 
submillimetre/millimetre wavelength range, inaccurate results could
lead to poor assumptions in galaxy number counts and galaxy evolution
models.  
The future ground-based, airborne and space observatories
exploring the FIR to millimetre wavelength regime, such as SOFIA,
ASTRO-F, Herschel, Planck and ALMA hold great expectations in the
investigation of extragalactic dust properties and will bring us many
steps closer ot the understanding of the physics of the ISM of
galaxies.


\begin{acknowledgements}
We would like to thank Ren\'e Gastaud, H\'el\`ene Roussel, Pierre Chanial and
Marc Sauvage for their expert advice on ISOCAM data reduction;
Andrew Baker, Thomas Stanke and Ute Lisenfeld for their assistance at IRAM
radio-telescope and their help on data reduction and Axel Weiss and Frank 
Bertoldi for their technical help;
Marc Sauvage, who provided us his M band image of \hen\ in advance of 
publication;
Leonardo Vanzi for his \nir\ images of \iizw;
Laurent Verstraete, who provided us his updated model of the PAH emission in 
the \dbp\ model;
Jean-Luc Starck, for useful discussion on data processing techniques. 
We also thank Ute Lisenfeld and Frank Israel for helpful scientific discussion.
We are very grateful to Fran\cc ois Boulanger and Eli Dwek
for invaluable insight on this subject which improved the scientific
quality of the paper. 
Finally, we thank the referee for his report, especially for 
his suggestion to consider the Akaike's Information Criterion.
\end{acknowledgements}


\appendix
\section{Degeneracies in dust modelling}

In this appendix, we demonstrate a well-known degeneracy between the 
energy source, and the grain properties,
which must be considered when modelling a dust SED.

The dust SED of a given region 
depends\textlist{\thetextlist~on the optical properties of the different types
         of grains, 
       \thetextlist~on the size distribution of these grains and 
       \thetextlist~on the shape and the intensity of the heating radiation 
field.}

\begin{figure*}[htbp]
  \centering
  \begin{tabular}{cc}
    \includegraphics[width=0.5\linewidth]{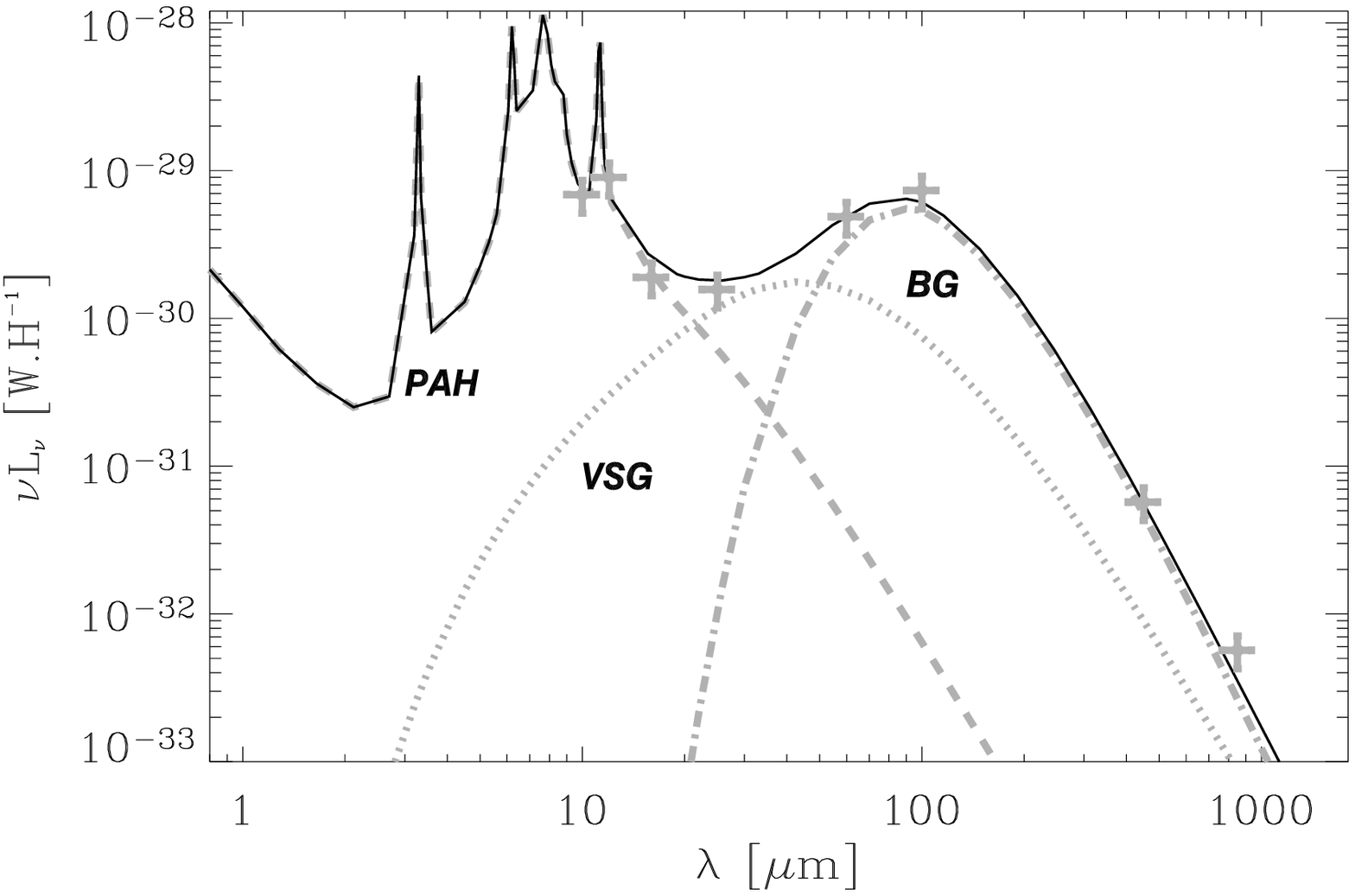}  & \hspace*{-0.6cm}
    \includegraphics[width=0.5\linewidth]{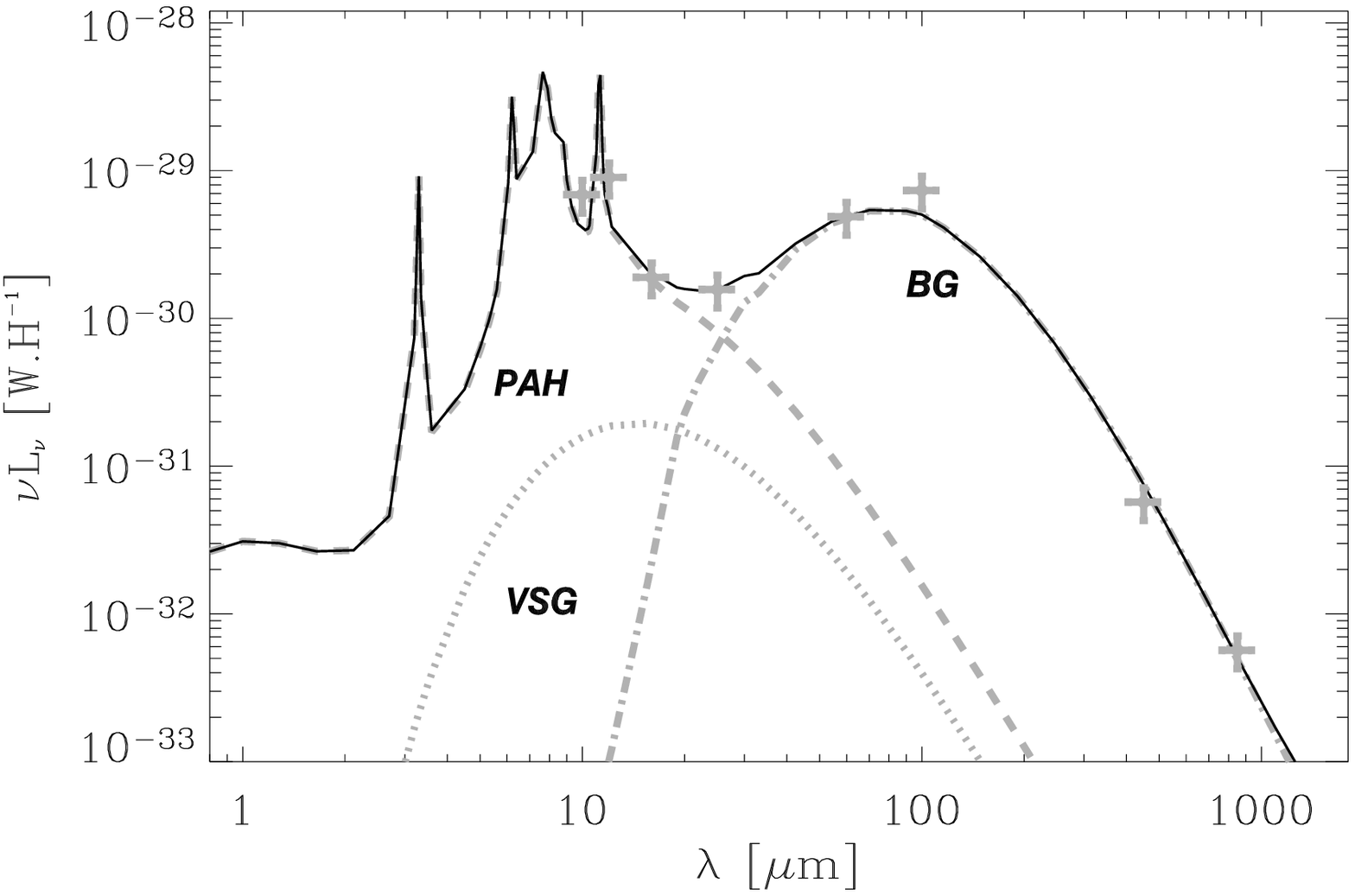}  \\
    \includegraphics[width=0.5\linewidth]{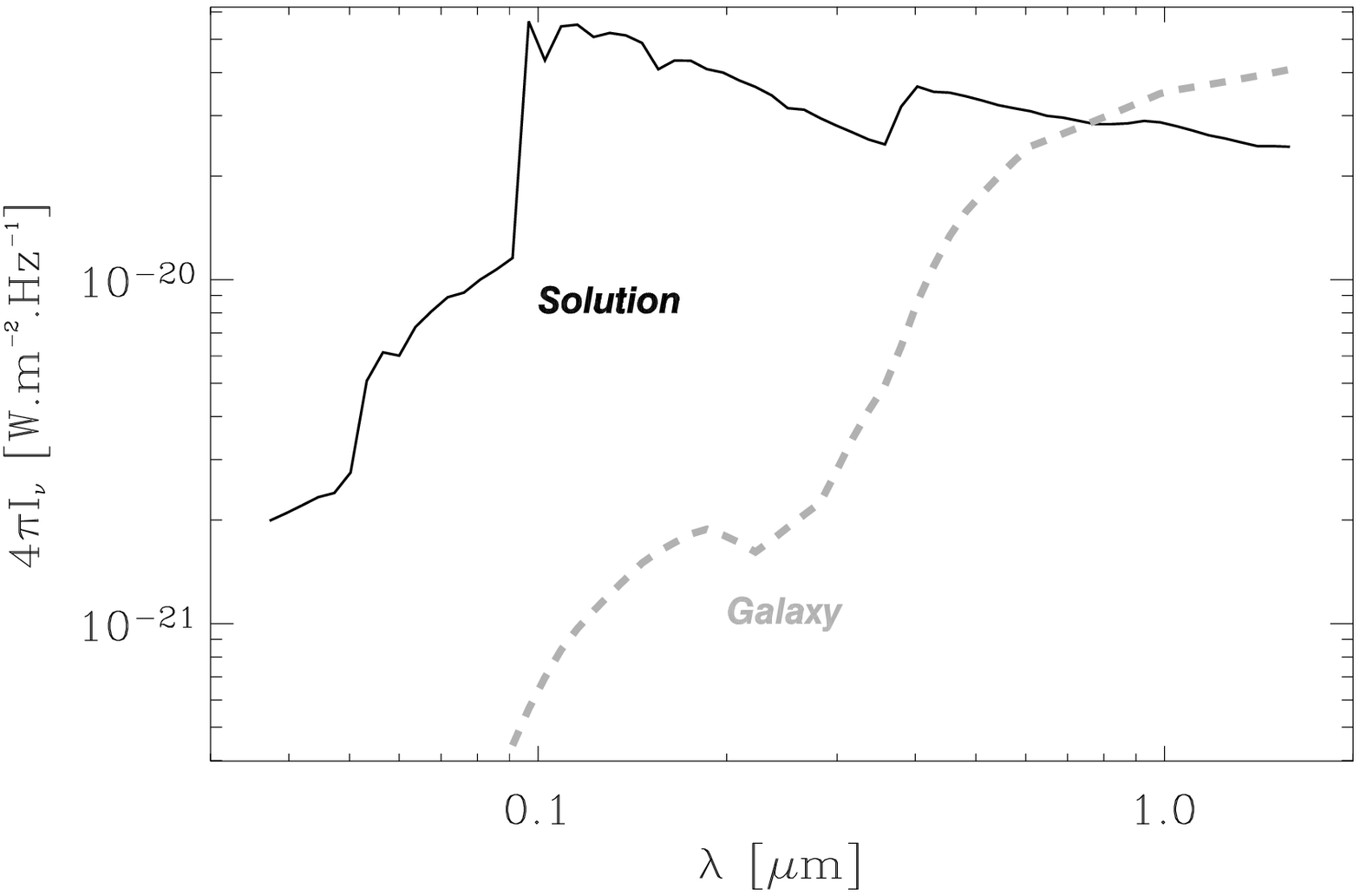} & \hspace*{-0.6cm}
    \includegraphics[width=0.5\linewidth]{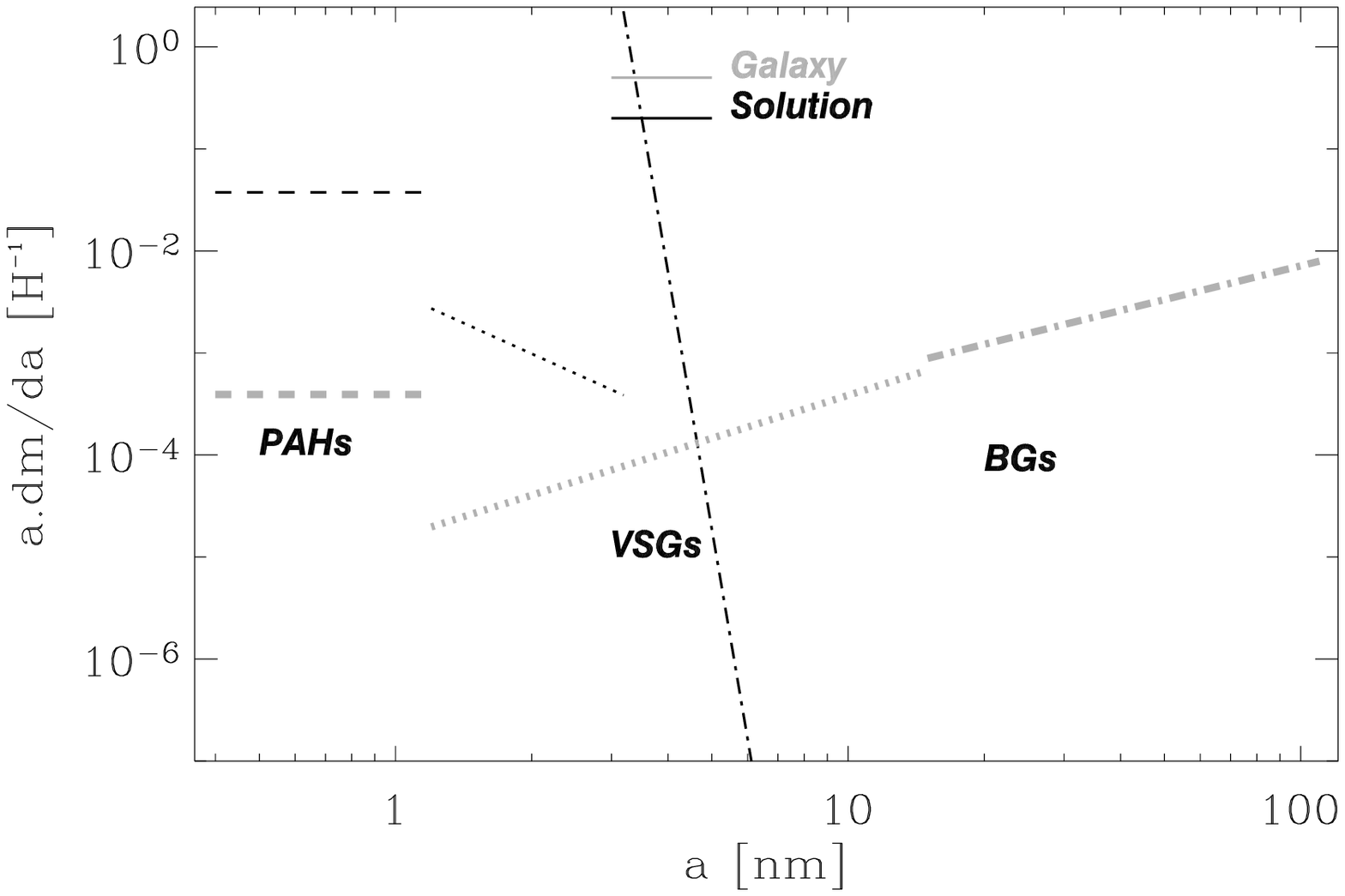} \\
  \end{tabular}
  \caption{The two top plots show the same fake observed dust SED (the grey 
           crosses) fitted with the \dbp\ model by varying two different sets 
           of parameters.
           The top-left plot has been fitted by varying the radiation field
           and keeping the Galactic size distribution.
           The top-right plot has been fitted by varying the size distribution
           and keeping the Galactic radiation field.
           The bottom-left plot shows the radiation field required for the
           top-left plot, compared to the Galactic one.
           The bottom-right plot shows the size distribution required for the
           top-right plot, compared to the Galactic one.
           }
  \label{fig:degeneracy}
\end{figure*}

Optical properties of the grains are based on laboratory experiments
and astrophysical measurements, it is not relevant to vary them in order to 
fit an observed SED. 
However, the radiation field depends on the environment.
The grains will experience a different radiation field in a starburst region
than in the diffuse ISM. 
The top-left plot of Fig.~\ref{fig:degeneracy} shows the fit of a fake
observed SED (the grey crosses) with the \citet{desert+90} model by varying 
the ISRF but keeping the Galactic grain size distribution.
The radiation field needed to fit this SED is the grey line on the bottom-left
plot on Fig.~\ref{fig:degeneracy}, compared to the Galactic one, in black.

The dust size distribution is also a set of parameters varying significantly 
in different environments.
Indeed, it is subject to fragmentation and erosion by shock waves, 
evaporation near strong radiation sources, or coagulation and accretion of
material in dense media.
The top-right plot of Fig.~\ref{fig:degeneracy} shows the same fake observed
SED than on the top-left plot, but fitted with the \citet{desert+90} model
by varying the size distribution and keeping the Galactic radiation field.
We get a good fit, in this case, too.
The bottom-right plot of Fig.~\ref{fig:degeneracy} shows the required size 
distribution to fit the SED, in grey, compared to the Galactic one, in black.

These two examples show that, without caution, we can interpret the
same SED with two different physical solutions.
We emphasize the need to constrain both the ISRF and the size distribution 
when modelling a dust SED.
Our modelling approach was aimed to achieve that goal.

\section{Peculiar optical properties to explain the submillimeter excess}
\label{app:agladze}

\begin{figure}[htbp]
  \centering
  \includegraphics[width=\linewidth]{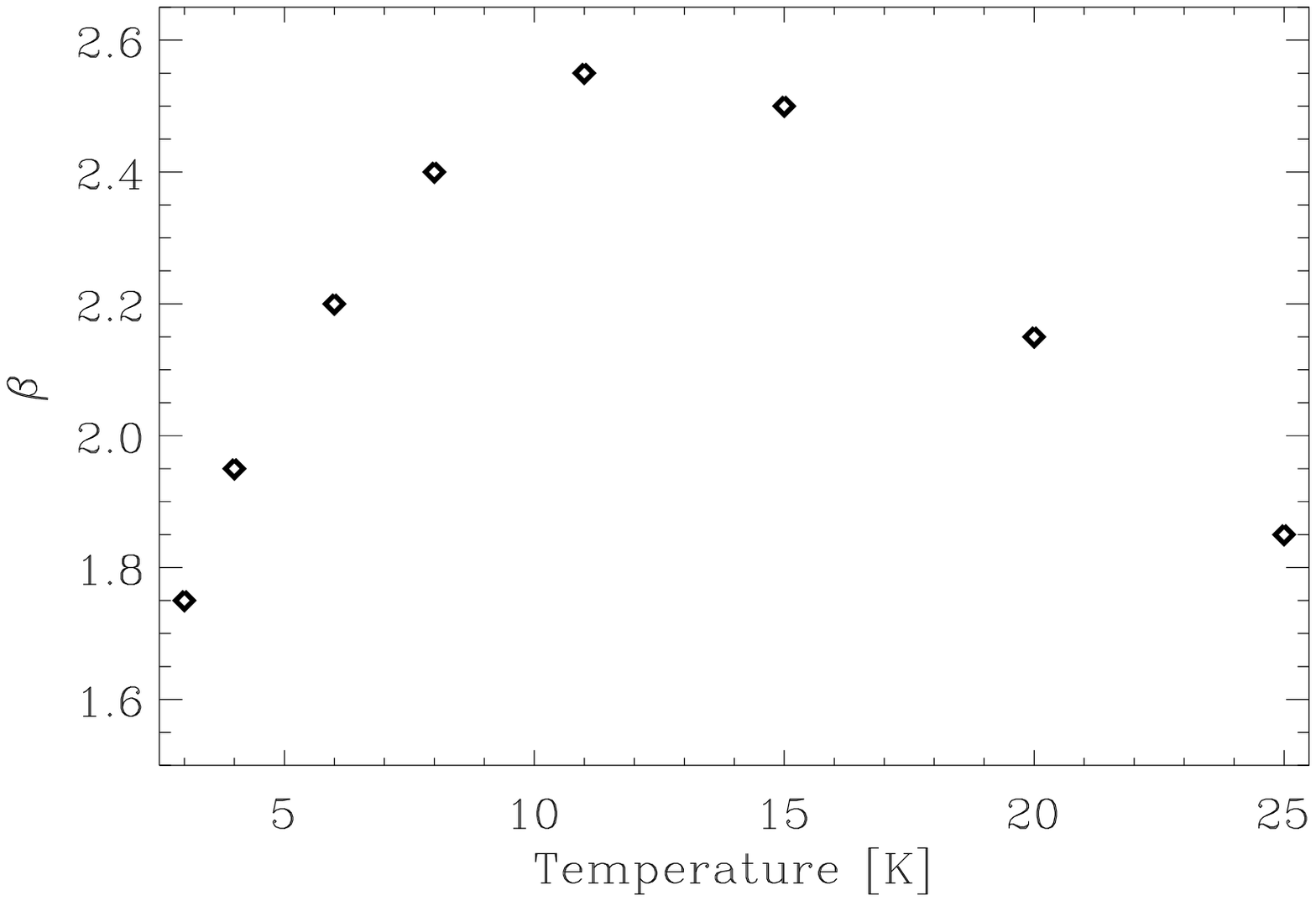}
  \caption{Variation of the emissivity index, $\beta$, as a function of the 
           temperature, $T$, for Mg$_2$SiO$_4$, as measured by 
           \citet{agladze+96}.}
  \label{fig:betagladze}
\end{figure}

\citet{agladze+96} measured the variation of the emissivity index, $\beta$, of
various silicate and carbonaceous materials with the temperature 
(Fig.~\ref{fig:betagladze}).
We have implemented their observations in order to explain the submillimeter
excess of our SEDs.

\begin{figure}[htbp]
  \centering
  \includegraphics[width=\linewidth]{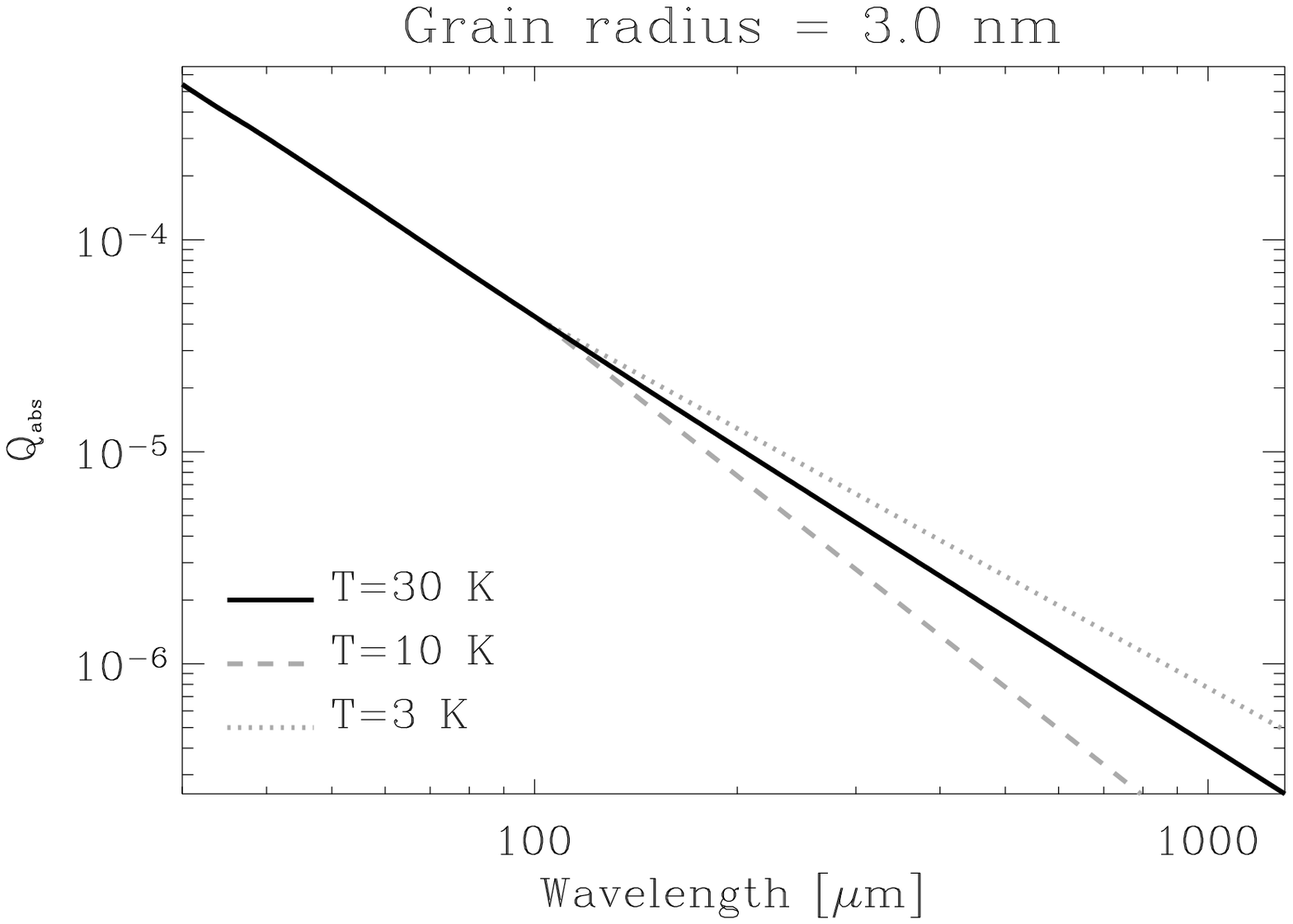}
  \caption{Modification of the optical properties of the BGs, following the
           measurements by \citet{agladze+96} shown in 
           Fig.~\ref{fig:betagladze}.}
  \label{fig:Qagladze}
\end{figure}

We modified the optical properties of the BGs, by constructing a temperature
dependent absorption efficiency $Q_\sms{abs}(\lambda,a,T)$ 
(Fig.~\ref{fig:Qagladze}).
In the standard \dbp\ model, the BGs are silicate grains and have an index
$\beta=2$, in the submillimeter regime.
The amorphous MgSiO$_3$ and Mg$_2$SiO$_4$ are the two species studied by
\citet{agladze+96} which are consistent with the BGs.
This modification can have a significant effect, if the contribution by 
cold BGs ($T_\sms{BG}\lesssim 25\,$K) is important.

For a given grain size, cold BGs can exist due to temperature fluctuations.
They cool down between two photon absorptions.
In our case, the monochromatic luminosity emitted by a BG of radius $a$ is:
\begin{equation}
  \mathcal{L}_\lambda (\lambda,a) = \int_0^\infty 4\pi B_\lambda (\lambda,T) 
                          \pi a^2 Q_\sms{abs}(\lambda,a,T) 
                          \frac{\ddiff P}{\ddiff T} 
                          \ddiff T,
\end{equation}
where $B_\lambda (\lambda,T)$ is the Planck function, and $\ddiff P/\ddiff T$ 
the temperature distribution.
Otherwise, BGs can be cold if they are large enough.
The total luminosity emitted by the BGs, with the number size distribution 
$f(a)$, is:
\begin{equation}
  L_\lambda(\lambda) = \int_0^\infty \mathcal{L}_\lambda (\lambda,a) f(a) 
                       \ddiff a.
\end{equation}

\begin{figure}[htbp]
  \centering
  \includegraphics[width=\linewidth]{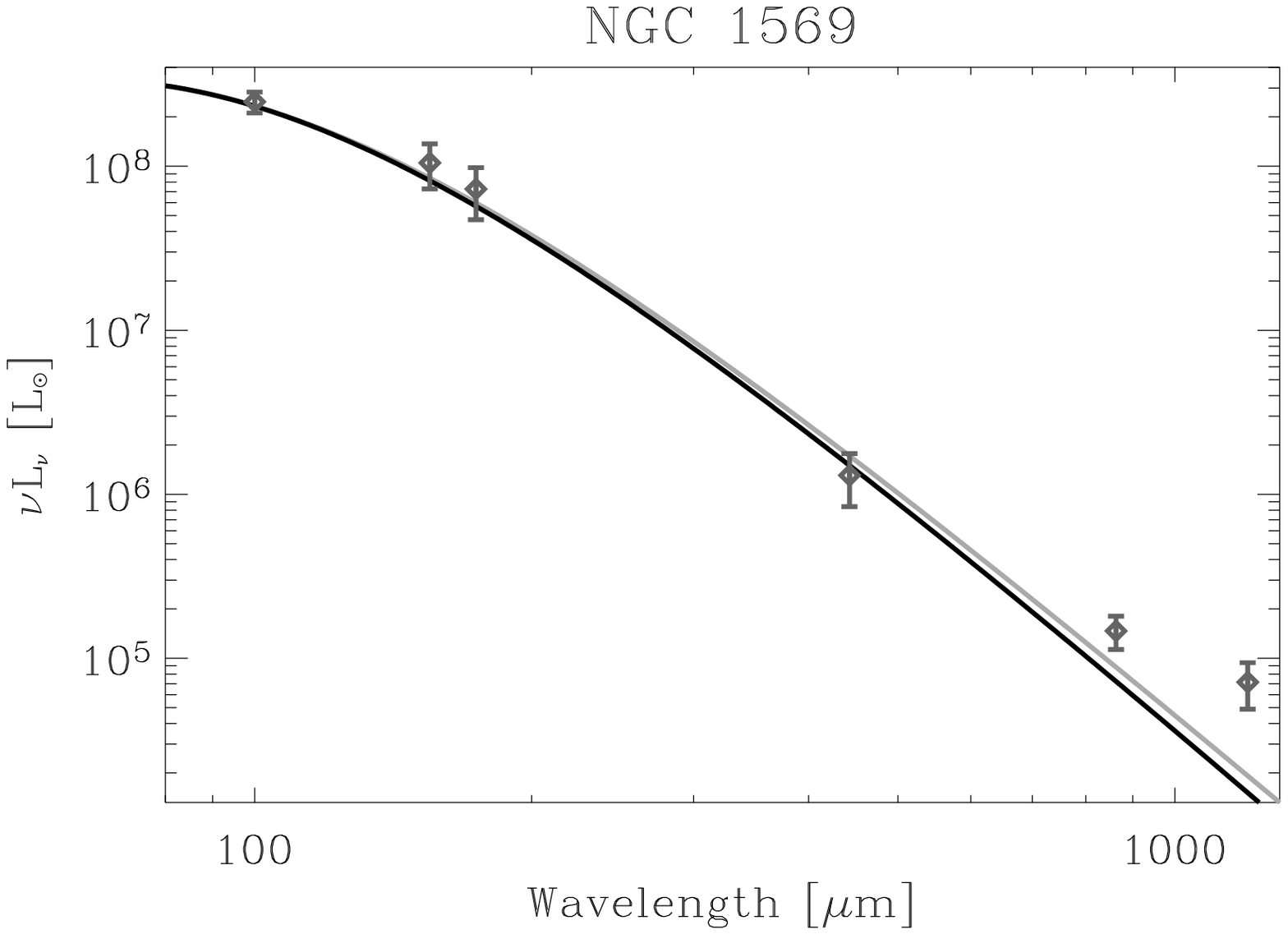}
  \caption{Fit of the long wavelength SED of \ngc{1569} (\papii), with
           the standard model (in black), and with the modified
           optical properties (in grey), using the temperature dependent
           $Q_\sms{abs}$.}
  \label{fig:SEDagladze}
\end{figure}

Among our four dwarf galaxies, the only detectable effect of these peculiar 
optical properties can be seen in \ngc{1569} (Fig.~\ref{fig:SEDagladze}).
For the three other SEDs, the two models give identical results.

Finally, we conclude that this effect is not able to reproduce the 
submillimeter excess.
This does not mean that we can exclude other peculiar, as yet unexplored, 
optical properties to explain this excess.


\bibliographystyle{aa}
\bibliography
{../BibTeX/article,../BibTeX/techreport,../BibTeX/book,../BibTeX/proceed}


\end{document}